\documentclass[aps,prl,reprint,longbibliography,amsmath,amssymb]{revtex4-1}

\usepackage{MnSymbol}
\usepackage{graphicx}
\usepackage{subfigure}
\usepackage{epstopdf}
\usepackage{hyperref}
\usepackage{bm}
\usepackage{color}
\usepackage[disable]{todonotes}
\usepackage{braket}
\usepackage{xcolor}
\usepackage{amsmath, bbm, verbatim}

\begin{document}

\title{Chiral Phonon Transport Induced by Topological Magnons}

\author{Even Thingstad}
\author{Akashdeep Kamra}
\author{Arne Brataas}
\author{Asle Sudb{\o}}
\affiliation{Center for Quantum Spintronics, Department of Physics, Norwegian University of Science and Technology, NO-7491 Trondheim, Norway}

\begin{abstract}
The plethora of recent discoveries in the field of topological electronic insulators has inspired a search for boson systems with similar properties. There are predictions that ferromagnets on a two-dimensional honeycomb lattice may host chiral edge magnons. In such systems, we theoretically study how magnons and phonons couple. We find topological {magnon}-polarons around the avoided crossings between phonons and topological magnons. Exploiting this feature along with our finding of {Rayleigh-like} edge phonons in armchair ribbons, we demonstrate the existence of chiral edge modes with a phononic character. We predict that these modes mediate a chirality in the coherent phonon response and suggest measuring this effect via elastic transducers. These findings reveal a possible approach towards heat management in future devices.
\end{abstract}

\maketitle

\textit{Introduction}.--- Topological electronic insulators~\cite{Haldane1988_Model,Kane2005_Z2,Kane2005_Quantum,Bernevig_2006,Hasan2010_Colloquium} 
are characterized by an insulating bulk with conducting `chiral' edge states. The unidirectional propagation of these chiral modes is `topologically protected' against defects at low temperatures when we can disregard inelastic scattering from phonons~\cite{Hasan2010_Colloquium}.
This has led to the development of a wide range of essential concepts, including
Majorana modes~\cite{Beenakker2013,Alicea2012,Mourik2012,Kane2008}, topological quantum computation~\cite{Nayak2008,DasSarma2005}, and chiral transport. Inspired by these findings, there has been an upsurge of efforts towards finding similar states in other systems~\cite{Jotzu2014_Experimental} with an emphasis on bosonic excitations~\cite{Lannebere2018_Link,Liu2017_Model, Owerre2016_First,Owerre2016_Topological, Kim2016_Realization, Owerre2017_Squeezed, Kim2017_Realizing}. There are predictions of topological magnons~\cite{Owerre2016_First, Owerre2016_Topological,Kim2016_Realization} in honeycomb ferromagnets with an engineered Dzyaloshinskii-Moriya interaction~\cite{Dzyaloshinskii1958_Thermodynamic,Moriya1960_Anisotropic} that induces the necessary band gap. In contrast to fermionic systems with Fermi energy within this band gap, the bulk is not necessarily insulating in bosonic systems~\cite{Ruckriegel2017_Bulk}.

The field of magnonics~\cite{Kruglyak2010,Chumak2015,Bauer2012,Uchida2010} focuses on pure spin transport mediated by magnons~\cite{Akhiezer1968}. It is possible to exploit the low-dissipation and wave-like nature of these excitations in information processing~\cite{Chumak2014,Ganzhorn2016}. The coherent pumping of chiral surface spin wave (Damon-Eshbach) modes induces cooling via incoherent magnon-phonon scattering~\cite{An2013}. Besides application oriented properties, the bosonic nature of magnons, combined with spintronic manipulation techniques~\cite{Saitoh2006,Chumak2015}, allows for intriguing physics~\cite{Sonin2010,Takei2014,Kamra2016A,Kamra2017}.  The coupling~\cite{Kittel1949} between magnons and phonons fundamentally differs from the electron-phonon interaction and results in a coherent hybridization of the modes~\cite{Kittel1958}, in addition to the temperature dependent incoherent effects~\cite{Uchida2011,An2013} discussed above. The direct influence of the hybridization between magnons and phonons, known as {magnon}-polarons~\cite{Kamra2015_Coherent,Kamra2014_Actuation}, has been observed in spin and energy transport in magnetic systems~\cite{Weiler2012,Dreher2012,Flebus2017,Kikkawa2016,Bozhko2017,Ruckriegel2014}.

In this Letter, we address the robustness of the topological magnons in a honeycomb ferromagnet~{\cite{Owerre2016_First, Owerre2016_Topological,Kim2016_Realization}} against their coupling with the lattice vibrations. In contrast to the case of electron-phonon coupling, where phonons can be disregarded at low temperatures, the magnon dispersion may undergo significant changes with new states emerging in the band gap~\cite{Bozhko2017,Ruckriegel2014}. {{We find that in the honeycomb ferromagnet with spins oriented orthogonal to the lattice plane,} only {transverse} phonon modes with {out-of-plane displacement} couple to spin. {To understand the eigenmodes}}, we evaluate and analyze the coupled spin and out-of-plane phonon modes for an infinitely large plane as well as for a finite ribbon geometry. We quantify the effect of the magneto-elastic coupling on the magnon Hall conductivity and find a non-monotonic dependence on the coupling strength. Our 
analysis of the finite ribbons 
shows that topological magnons hybridize with bulk phonons around the avoided crossings in their coupled dispersion, forming {magnon}-polarons with topological chiral properties. Hence, while their edge localization is weakened, the magneto-elastic coupling does not completely remove the topological magnons. Furthermore, we find that armchair edges support {Rayleigh-like} edge phonon modes in {sharp} contrast to the zigzag edges. When topological magnons hybridize with these edge phonons, edge {magnon}-polarons with almost undiminished chirality are formed. We suggest a setup which utilizes this induced chirality in coherent phonon transport. Such systems enable the observation of the topological physics and serve as a prototype for a unidirectional heat pump. {This offers a highly feasible alternative to producing topological phonon diodes \cite{RevModPhys.84.1045,Maldovan_Nature_2013,Zhang_Topo_Phonon_2016}.}

\textit{Model}.---We consider a ferromagnetic material with localized spins on a two-dimensional honeycomb lattice, allow for out-of-plane vibrations of the lattice sites, and assume there is magneto-elastic coupling. This system can be modelled by a Hamiltonian of the form \(H=H_m + H_\text{ph} + H_\text{me}\), where \(H_m\) is the magnetic Hamiltonian, \(H_\text{ph}\) describes the phonons, and \(H_\text{me}\) represents the magneto-elastic coupling. 

\begin{figure}
    \centering
    \includegraphics[width=0.95\columnwidth]{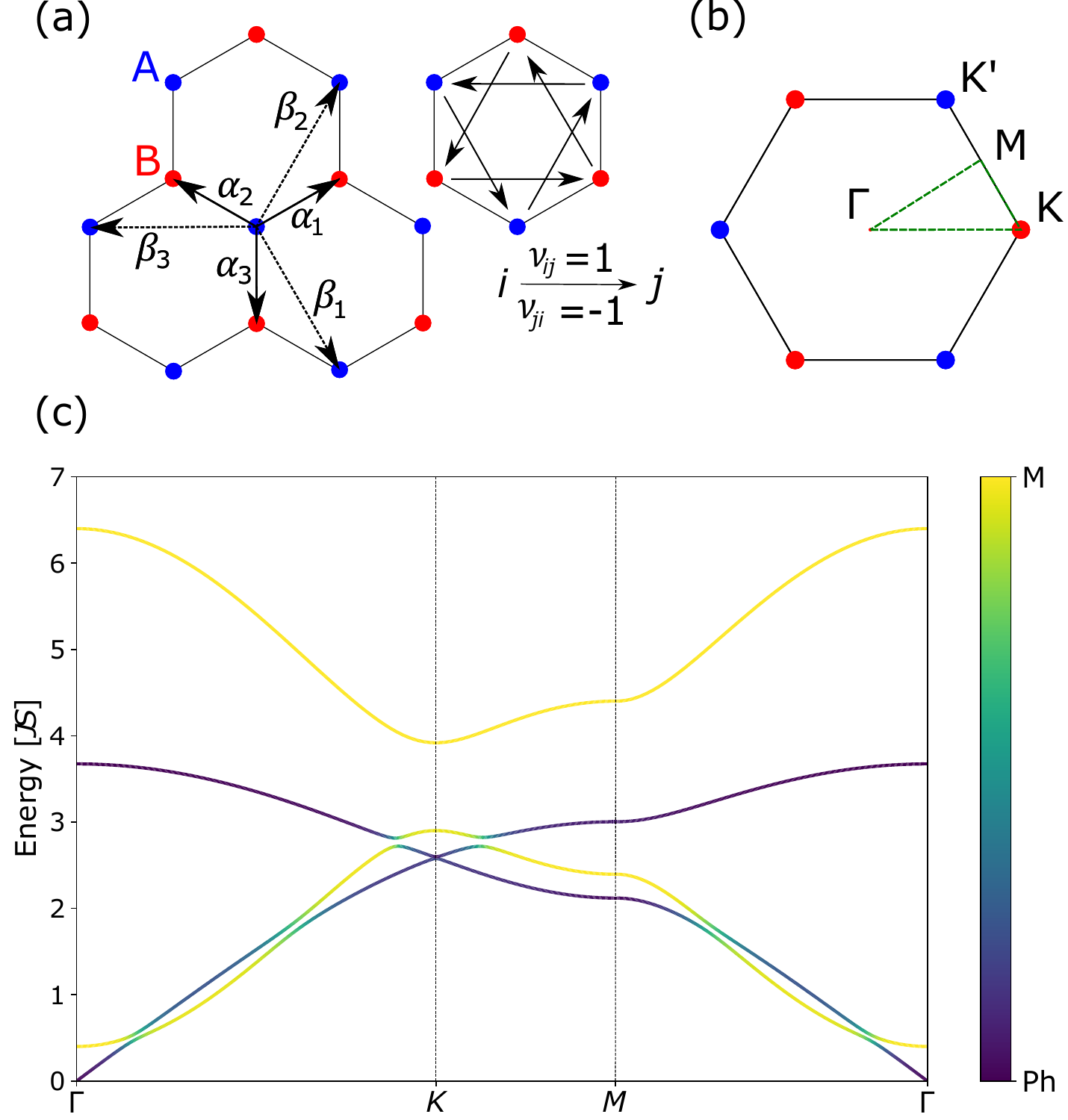}
    \caption{(a) Lattice geometry showing the nearest neighbour vectors \(\pmb{\alpha}\), next-to-nearest neighbour vectors \(\pmb{\beta}\), and the Haldane sign \(\nu_{ij}=\pm 1\). (b) The first Brillouin zone in reciprocal space, including the paths along which we plot the dispersion relation in figure (c). The parameter values used are \(\mathcal{D}=0.1 J\), \(\mathcal{B}=0.4JS\), \(\sqrt{C/m}=1.5JS\), and rescaled coupling strength \(\tilde{\kappa}=0.03\) (see main text). The magnon (yellow) and phonon (purple) character of the modes is indicated with colors. The modes are significantly affected by the magneto-elastic coupling only close to avoided crossings.}
    \label{fig_1}
\end{figure}

The  Hamiltonian we consider is 
inspired by the Haldane model \cite{Haldane1988_Model} 
given by 
\cite{Owerre2016_First, Owerre2016_Topological, Kim2016_Realization}
\begin{equation}
H_m = -J \sum_{\langle i,j \rangle} \mathbf{S}_i \cdot \mathbf{S}_j + \mathcal{D} \sum_{\llangle i,j \rrangle} \nu_{ij} \hat{z} \cdot \mathbf{S}_i \times \mathbf{S}_j - \mathcal{B} \sum_i S_i^z.
\end{equation}
\noindent The first term describes the ferromagnetic exchange coupling between nearest neighbour sites, while the second accounts for the Dzyaloshinskii-Moriya interaction \cite{Dzyaloshinskii1958_Thermodynamic, Moriya1960_Anisotropic} between next-to-nearest neighbours {\footnote{The demagnetization energy is disregarded since it only causes minor shifts in the dispersion 
~\cite{Kamra2017}.}}. The Haldane sign  \(\nu_{ij} = \pm 1\) depends on the relative orientation of the next-to-nearest neighbours as shown in Fig. \ref{fig_1}(a), and is the root of
non-trivial topological properties. We let the nearest neighbour distance be \(d\) and the next-to-nearest neighbour distance be \(a\).
Refs. {\cite{Owerre2016_Topological, Kim2016_Realization}} discuss the dispersion relation and Berry curvature of this spin model in linear spin wave theory.

For the phonon Hamiltonian, we consider only the out-of-plane degrees of freedom since only these modes couple to the spin to lowest order in the linear spin wave expansion. We assume nearest-neighbour interactions with elastic constant \(C\), let the mass associated with the spins on the lattice sites be \(m\), {and disregard substrate coupling}. Introducing \(S_k = \sum_\beta \cos ( \mathbf{k} \cdot \pmb{\beta}) \),
where the sum is over the three next-to-nearest neighbour vectors \(\pmb{\beta}\) of Fig. \ref{fig_1}(a), we obtain the dispersion relation
\begin{equation}
\omega^\text{ph}_\pm (\mathbf{k}) 
= \sqrt{\frac{C}{m}} \sqrt{3 \pm \sqrt{3 + 2 S_k}} 
\end{equation}
\noindent for the free phonon modes. 

Motivated by the continuum limit description~\cite{Kittel1949,Kittel1958}, we write down the lattice magneto-elastic coupling to linear order in the magnon amplitude, obtaining
\begin{equation}
H_\text{me} = \kappa \sum_D \sum_{i\in D} \sum_{ \pmb{\alpha}_D} {\mathbf{S}_i \cdot  \pmb{\alpha}_D} \; (u_{i}^z - u_{i + \pmb{\alpha}_D}^z )
\end{equation}

\noindent where \(\kappa\) parametrizes the strength of the magnon-phonon coupling,  \( \sum_D \) denotes the sum over sublattices,  \( \sum_{i\in D} \) is the sum over the lattice sites on the \(D\) sublattice, and \(\pmb{\alpha}_D\) are the corresponding nearest neighbour vectors. The out-of-plane deviation for lattice site \(i\) is denoted by \(u_{i}^z\).

\textit{Bulk spectrum}.---We introduce the Holstein-Primakoff representation of spins and use linear spin wave theory in the spin- and magneto-elastic terms~\cite{Akhiezer1968}. Within the rotating wave approximation \cite{Scully1997_Quantum}, the resulting Hamiltonian describing the phonon and magnon modes of the system is obtained as \(H=\ \sum_k \psi_k^\dagger M_k \psi_k\), where
\(\psi_k^\dagger = (a_k^\dagger, b_k^\dagger, c_{k-}^\dagger, c_{k+}^\dagger)\). 
Here, \(a_k\) and \(b_k\) are annihilation operators for the sublattice magnon modes on the \(A\) and \(B\) sublattices, while \(c_{k\pm}\) are the annihilation operators for the phonon modes. 
The matrix \(M_k\) takes the form

\begin{equation}
M_k = \begin{pmatrix}
A + h^z & h^- & g_{A-} & g_{A+} \\
h^+ & A-h^z & g_{B-} & g_{B+} \\
g_{A-}^* & g_{B-}^* & \omega_{k-}^\text{ph} & 0 \\
g_{A+}^* & g_{B+}^* & 0 & \omega_{k+}^\text{ph} \\
\end{pmatrix},
\end{equation}

\noindent where \(A=3JS+\mathcal{B}\), \(h^z(\mathbf{k}) = 2\mathcal{D} S \sum_\beta \sin (\mathbf{k} \cdot \pmb{\beta})\), \(h^-(\mathbf{k}) = -JS \sum_\alpha \exp ( -i \mathbf{k} \cdot \pmb{\alpha} )\), and \(h^+ = (h^-)^*\). The coupling between the \(D\)-sublattice magnons and the phonon branch \(\pm\) is captured by \(g_{D\pm}\), which is proportional to the dimensionless coupling strength \(\tilde{\kappa} = (\kappa d/JS) \sqrt[4]{\hbar^2 S^2/16m^2 (C/m)}\). The spectrum obtained by diagonalizing this matrix is plotted in Fig. \ref{fig_1}(c) along the paths displayed in Fig. \ref{fig_1}(b).

\textit{Hall conductivity}.---The topological nature of the spin model is manifested in the magnon Hall conductivity that arises because of the time-reversal symmetry breaking 
caused by the Dzyaloshinskii-Moriya interaction. 


The spin current operator $J_\gamma$ may be found from  a continuity equation or magnon group velocity approach~\cite{Nakata2017_Magnonic}, both yielding

\begin{equation}
J_\gamma = \sum_k 
\begin{pmatrix} a_k^\dagger & b_k^\dagger \end{pmatrix} \left( \frac{\partial H_m(\mathbf{k})}{\partial k_\gamma} \right) 
\begin{pmatrix} a_k \\ b_k \end{pmatrix}
\end{equation}

\noindent  along the Cartesian direction \(\gamma\). Here, \(H_m(\mathbf{k} )\) is the matrix representation of the magnon Hamiltonian. Assuming we apply a weak in-plane magnetic field gradient \(\nabla \mathcal{B}\), we are interested in the current \(\mathbf{j} = \sigma \nabla \mathcal{B}\), which is determined by the conductivity tensor \(\sigma\)~\cite{Nakata2017_Magnonic}. The Hall conductivity can be calculated using the Kubo formula, giving  

\begin{equation}
\sigma_{xy} =  \sum_k \sum_{\alpha, \beta \neq \alpha} n_B(E_{k\alpha}) C_{\alpha\beta} (\mathbf{k}), 
\label{eq_hallConductivity}
\end{equation}

\noindent where \(E_{k\alpha}\) is the energy eigenvalue of band \(\alpha\) and \(n_B(E_{k\alpha})\) is the corresponding Bose factor. The curvature-tensor \(C_{\alpha\beta}\) is given by 

\begin{equation}
C_{\alpha\beta} (\mathbf{k}) = i  \frac{J^{\alpha\beta}_y (\mathbf{k}) J^{\beta\alpha}_x(\mathbf{k}) - J^{\alpha\beta}_x (\mathbf{k}) J^{\beta\alpha}_y(\mathbf{k})}{(E_{k\alpha} - E_{k\beta})^2},
\end{equation}

\noindent where \( (\alpha,\beta) \) are band-indices, and \(J^{\alpha\beta}_{\gamma}(\mathbf{k})\) are the energy eigenstate matrix elements of the current operator at quasimomentum \(\mathbf{k}\). Disregarding the magneto-elastic coupling, the  band-curvature \(C_\alpha = \sum_{\beta \neq \alpha} C_{\alpha\beta}\) can be identified as the Berry curvature.

\begin{figure}
    \begin{center}
    \includegraphics[width=1\columnwidth]{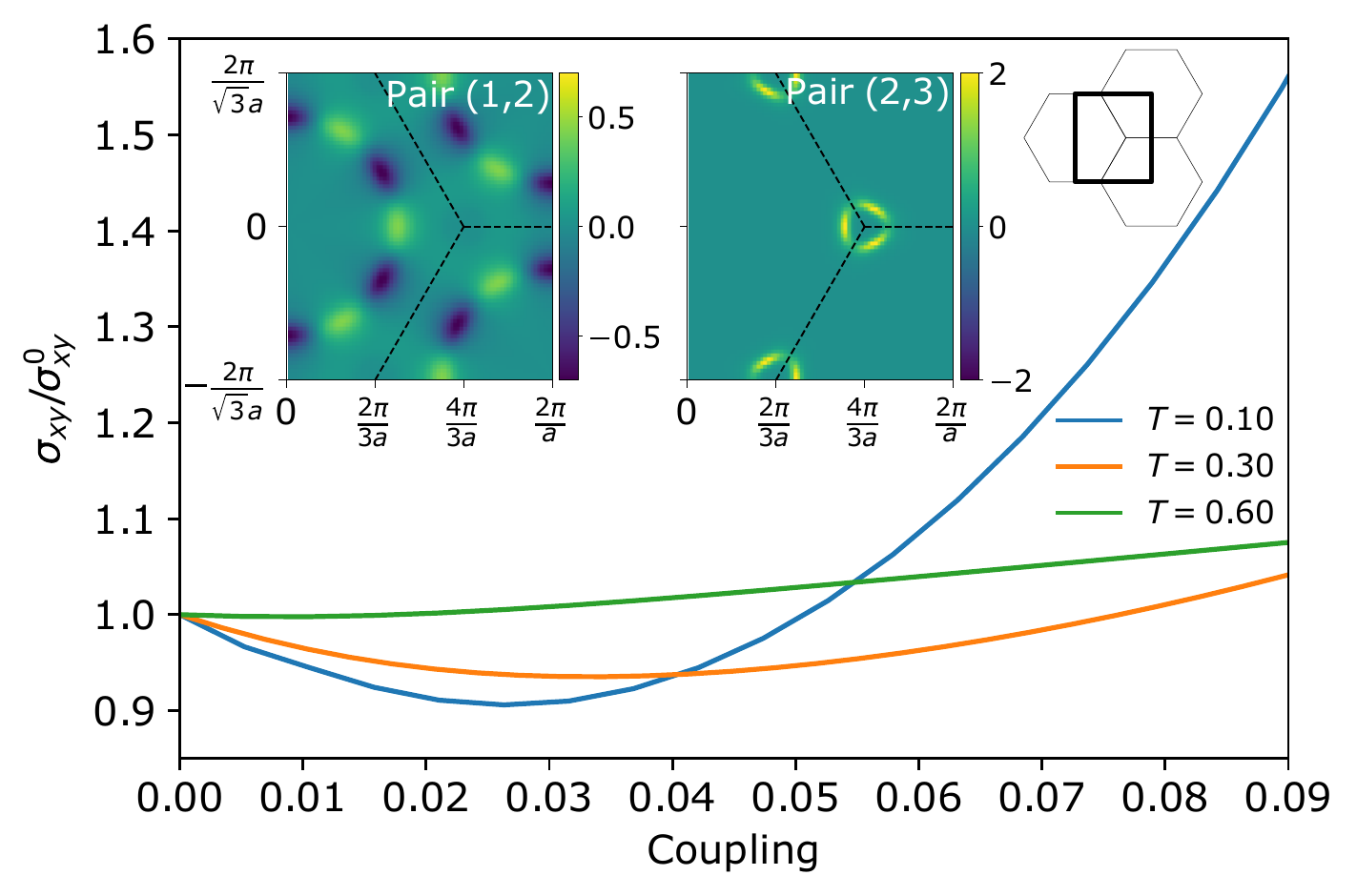}
    \end{center}
    \caption{Dependence of the Hall conductivity on the magneto-elastic coupling strength \(\tilde{\kappa}\) for parameter values \(\mathcal{D}=0.1 J\), \(\mathcal{B}=0.4JS\), and \(\sqrt{C/m}=1.5 JS\) for different temperatures \(T\) in units of \(JS\). The insets show the quasimomentum dependence of the curvature-tensor at \(\tilde{\kappa}=0.03\) for band-pairs (1,2) and (2,3), where the bands are labelled according to their energy and band 1 is the lowest band. The dominant contribution in these band-pairs comes from the regions with avoided crossings of the respective bands. }
    \label{fig_2}
\end{figure}

Expressing the sublattice magnon operators in terms of the eigenmode operators, one may identify the current matrix elements \(J_\gamma^{\alpha\beta}\) and integrate the curvature over the Brillouin zone to obtain the Hall conductivity. We are particularly interested in the effect of the magneto-elastic coupling, and therefore present the dependence of the Hall conductivity on the dimensionless coupling \(\tilde{\kappa}\) in Fig \ref{fig_2}. 

To understand this dependence, we consider the curvature-tensor \(C_{\alpha\beta}\). When the bands \(\alpha\) and \(\beta\) both have a predominant magnon content, the topological nature of the underlying magnons gives a finite curvature. This magnon curvature is largest close to the Dirac points~\cite{Owerre2016_Topological, Kim2016_Realization}. Close to an avoided crossing, the magneto-elastic coupling changes the spectrum and causes transfer of band-curvature between the relevant bands \(\alpha\) and \(\beta\). The latter can be seen by plotting the curvature-tensor element \(C_{\alpha\beta}\) for the band-pairs with avoided crossings, as shown in the insets of Fig \ref{fig_2}. The resulting change in Hall conductivity is given by these curvature-tensor elements weighted with the difference between the Bose factors of the relevant bands. This follows from the anti-symmetry property of the curvature-tensor. The two band-pairs in the insets contribute oppositely to the Hall conductivity, and the competition between their curvature transfer explains the non-monotonic behaviour of the Hall conductivity.

\textit{Ribbon geometry and coherent transport}.---Due to the topological nature of the magnon model under consideration and the bulk-boundary correspondence, there are gapless magnon edge states in a finite sample \cite{Hasan2010_Colloquium, Owerre2016_First, Owerre2016_Topological, Kim2016_Realization}. {Considering an armchair ribbon with finite {width}, the one-dimensional projection of the energy spectrum is plotted in Fig. \ref{fig_3}. The corresponding spectrum for the zigzag edge ribbon is given in the Supplemental Material \cite{suppMat}, where also Refs. \cite{Kittel1987,Kino1987,Maradudin1991,Bernevig2013,Ruello2015} appear.} Magnon and phonon modes hybridize in regions with an avoided crossing. When the upper phonon band lies within the bandgap of the pure magnon spectrum, there are modes with a mixed content of chiral magnon edge states and phonons. Although the spectra look qualitatively similar, there is a crucial distinction between the two cases. For the zigzag edge configuration, all the phonon modes are delocalized throughout the sample, {while the armchair edges host {``Rayleigh-like''}  edge phonon modes. In direct analogy with Rayleigh modes on the surface of a three-dimensional material, the localization length of these modes is {directly proportional to their wavelength}, as shown in the Supplemental Material \cite{suppMat}.
These edge phonon modes are {supported by}} the half-hexagon protrusions of the armchair edge {that can} pivot around the bonds parallel to the edges connecting these protrusions, see Fig. \ref{fig_3}. No such parallel bonds exist for the zigzag edge. 


The Hall conductivity is a hallmark of topological electronic properties and motivates a similar role for the Hall conductivity mediated by topological magnons. However, in contrast to electrons, the bosonic nature of the magnons results in the lack of a general proportionality between the magnon Hall conductivity and the Chern number~\cite{Nakata2017_Magnonic}. Furthermore, the observation of a magnon planar Hall effect~\cite{Liu2017} in a cubic, non-topological magnet suggests that this Hall conductivity may not be regarded as a smoking-gun signature for topological properties. Thus, we suggest a complementary approach to observe the topological nature of the underlying magnons by elastically probing the chirality of the magneto-elastic hybrid modes. 

\begin{figure}
    \centering
    \includegraphics[width=0.9\columnwidth]{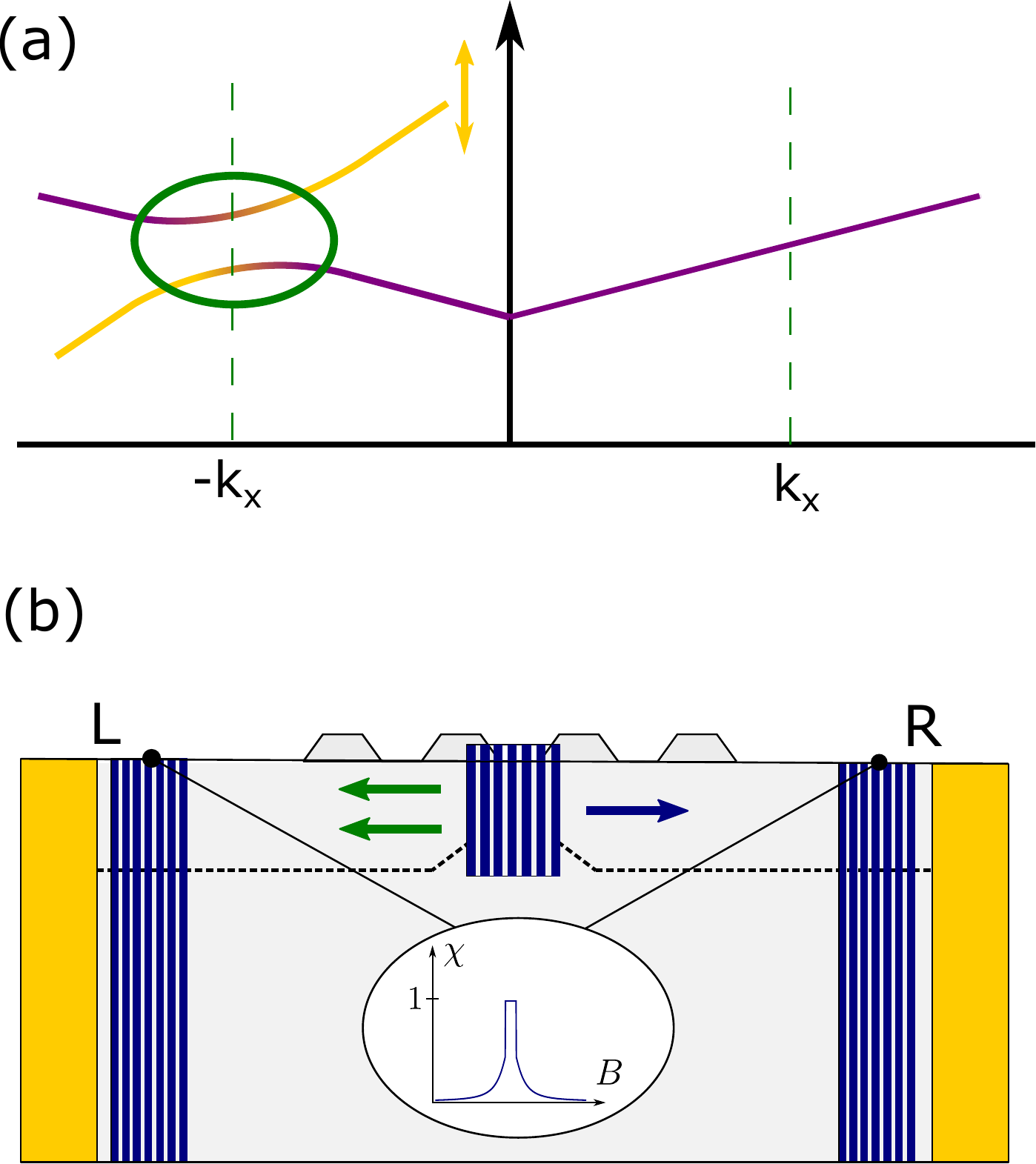}
    \caption{{(a) Schematic spectrum for the {coupled} Rayleigh-like edge phonons and {the} topological edge magnon on the upper {armchair sample} edge. {The phonon at quasimomentum \(-k_x\) hybridizes with the chiral magnon, while the phonon at quasimomentum \(+k_x\) is unaffected due to the lack of a magnon with matching wavevector at this edge.} At the avoided crossing, there is propagation direction reversal for the modes with a phononic content. {The color of the dispersion represents its nature with purple representing phononic and yellow magnonic character}.} (b) Proposed experimental setup for detecting coherent chiral transport through excitation of phononic modes. Elastic energy is injected {in the sample middle on the upper armchair edge and detected at the left (L) and right (R) edges using wavevector and frequency resolved elastic transducers (purple). By exciting modes at the avoided crossing in (a), only the elastic excitations at one of the two quasimomenta \(\pm k_x\) are converted into hybridized modes (green arrow{s}). This gives a chiral response, and the chirality is peaked when the wavevector of the avoided crossing coincides with the fixed wavevector of the transducer.} }
    \label{fig_4}
\end{figure}

We propose to observe coherent chiral phonon propagation in the experimental setup of Fig.~\ref{fig_4}(b) {by utilizing the edge modes, as depicted schematically in Fig. \ref{fig_4}(a)~\footnote{The inset of Fig. \ref{fig_3}(b) shows two phonon modes. One is irrelevant since it is localized on the opposite edge.}, on the upper armchair edge of the sample.} {Taking inspiration from previous related experiments~\cite{Weiler2012,Gowtham2015}, we suggest to inject elastic energy into} the sample middle {at the upper edge} using a nano-scale variant of the interdigital transducer design~\cite{Datta1986, Mamishev2004}, {elaborated further in the Supplemental Material \cite{suppMat}}. For a given transducer {design}, modes are excited with fixed wavevectors \(\pm k_x\) and a tunable frequency. {Similar} transducers can be used to detect {the} elastic response \(p_{L/R}\) {on} the left (L) and right (R) edges of the sample. Here, \(p_{L/R} \) is the {elastic power detected} at the transducers.


{{Figure \ref{fig_4}(a) schematically depicts the dispersion for the magnetoelastic modes localized on an armchair edge. Disregarding magnetoelastic coupling, the edge hosts two counterpropagating Rayleigh-like edge phonons and a single chiral edge magnon. There is thus no chirality in the phononic response. Due to magnetoelastic coupling, the Rayleigh-like phonon with wavevector $-k_x$ hybridizes with the chiral magnon to form a magnon-polaron while the other phonon remains unchanged. This breaks the symmetry between the counterpropagating phononic modes and} the result is non-zero chirality \(\chi=(p_R - p_L)/(p_R + p_L)\). Furthermore, as shown {in} Fig. \ref{fig_4}(a), the hybridization with the magnon mode reverses the {group velocity direction of the participating phonon mode}. In principle, this gives {perfectly} chiral phonon transport.}

{
The wavevector location of the avoided crossing can be tuned {via the Zeeman shift in the magnon dispersion}. Performing a frequency integrated measurement over an energy range of the same order as the magneto-elastic coupling, one obtains a peaked chirality when the  magnetic field is such that the wavevector of the avoided crossing coincides with the wavevector of the transducer, obtaining a chirality as shown in Fig. \ref{fig_4}(b). Performing a similar transport experiment {on the zigzag edge does} not give chiral phonon transport since the delocalized phonons {hybridize with counterpropagating magnons on both the edges, thereby destroying the overall chirality}. In addition, the size of the avoided crossing {is} smaller due to the smaller overlap with the localized chiral magnon. {The} armchair edge is therefore crucial {for obtaining the} chirality.}

\begin{figure}
    \centering
    \includegraphics[width=1\columnwidth]{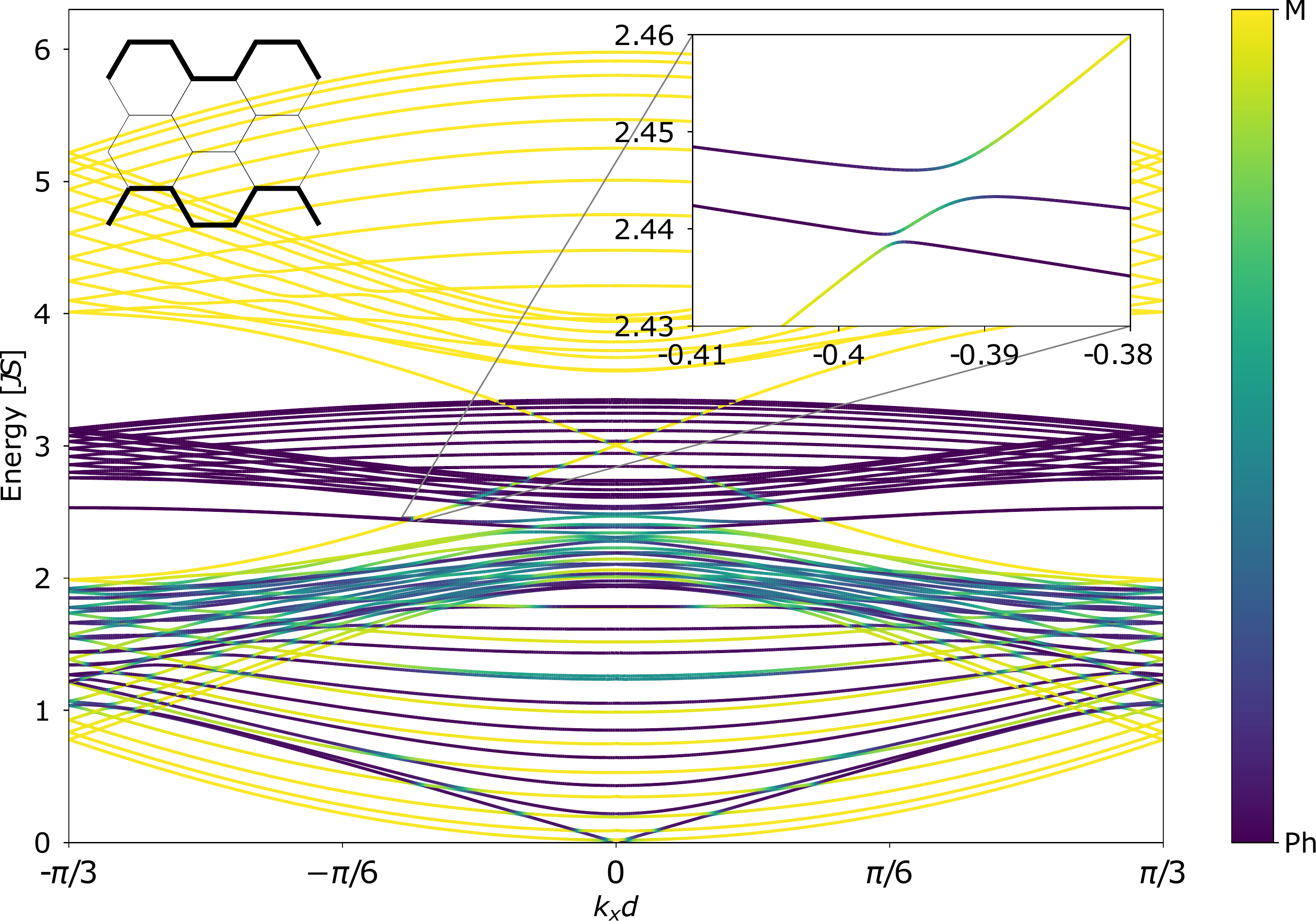}
    \caption{One-dimensional projection of the dispersion relation for the magnetoelastic modes on a honeycomb ribbon with armchair edges. In addition to the bulk bands, there are two topological edge magnon states crossing the magnon band gap, as well as  Rayleigh-like edge phonons. 
    The inset shows the avoided crossing of a topological magnon edge mode with the two quasi-degenerate edge phonon modes.  The parameter values are \(\mathcal{B}=0\), \(\sqrt{C/m} = 1.37 JS\), \(\mathcal{D}=0.1J\) and \(\tilde{\kappa}=0.03\). }
    \label{fig_3}
\end{figure}

\textit{Summary}.--- We have examined the robustness of topological magnons in a honeycomb ferromagnet against their interaction with phonons. 
Their topological properties, albeit weakened, survive the magneto-elastic coupling. The magnon Hall conductivity of the system is found to depend on the magneto-elastic coupling strength in a non-monotonic, temperature-sensitive manner. Exploiting the {Rayleigh-like} edge phonons in armchair ribbons, we predict the existence of topological {magnon}-polarons confined to the boundary.  We have suggested an experimental setup capable of probing the chiral nature of the topological {magnon}-polarons by elastic means, which thus serves as a platform for chiral coherent phononic transport.

\textit{Acknowledgments}.---We acknowledge support from the Research Council of Norway Grant Nos. 262633 ``Center of Excellence on Quantum Spintronics'', and 250985, ``Fundamentals of Low-dissipative Topological Matter''. 

\bibliography{main_thirdVersion}

\begin{thebibliography}{63}%
\makeatletter
\providecommand \@ifxundefined [1]{%
 \@ifx{#1\undefined}
}%
\providecommand \@ifnum [1]{%
 \ifnum #1\expandafter \@firstoftwo
 \else \expandafter \@secondoftwo
 \fi
}%
\providecommand \@ifx [1]{%
 \ifx #1\expandafter \@firstoftwo
 \else \expandafter \@secondoftwo
 \fi
}%
\providecommand \natexlab [1]{#1}%
\providecommand \enquote  [1]{``#1''}%
\providecommand \bibnamefont  [1]{#1}%
\providecommand \bibfnamefont [1]{#1}%
\providecommand \citenamefont [1]{#1}%
\providecommand \href@noop [0]{\@secondoftwo}%
\providecommand \href [0]{\begingroup \@sanitize@url \@href}%
\providecommand \@href[1]{\@@startlink{#1}\@@href}%
\providecommand \@@href[1]{\endgroup#1\@@endlink}%
\providecommand \@sanitize@url [0]{\catcode `\\12\catcode `\$12\catcode
  `\&12\catcode `\#12\catcode `\^12\catcode `\_12\catcode `\%12\relax}%
\providecommand \@@startlink[1]{}%
\providecommand \@@endlink[0]{}%
\providecommand \url  [0]{\begingroup\@sanitize@url \@url }%
\providecommand \@url [1]{\endgroup\@href {#1}{\urlprefix }}%
\providecommand \urlprefix  [0]{URL }%
\providecommand \Eprint [0]{\href }%
\providecommand \doibase [0]{http://dx.doi.org/}%
\providecommand \selectlanguage [0]{\@gobble}%
\providecommand \bibinfo  [0]{\@secondoftwo}%
\providecommand \bibfield  [0]{\@secondoftwo}%
\providecommand \translation [1]{[#1]}%
\providecommand \BibitemOpen [0]{}%
\providecommand \bibitemStop [0]{}%
\providecommand \bibitemNoStop [0]{.\EOS\space}%
\providecommand \EOS [0]{\spacefactor3000\relax}%
\providecommand \BibitemShut  [1]{\csname bibitem#1\endcsname}%
\let\auto@bib@innerbib\@empty
\bibitem [{\citenamefont {Haldane}(1988)}]{Haldane1988_Model}%
  \BibitemOpen
  \bibfield  {author} {\bibinfo {author} {\bibfnamefont {F.~D.~M.}\
  \bibnamefont {Haldane}},\ }\bibfield  {title} {\enquote {\bibinfo {title}
  {Model for a quantum hall effect without landau levels: Condensed-matter
  realization of the "parity anomaly"},}\ }\href {\doibase
  10.1103/PhysRevLett.61.2015} {\bibfield  {journal} {\bibinfo  {journal}
  {Phys. Rev. Lett.}\ }\textbf {\bibinfo {volume} {61}},\ \bibinfo {pages}
  {2015--2018} (\bibinfo {year} {1988})}\BibitemShut {NoStop}%
\bibitem [{\citenamefont {Kane}\ and\ \citenamefont
  {Mele}(2005{\natexlab{a}})}]{Kane2005_Z2}%
  \BibitemOpen
  \bibfield  {author} {\bibinfo {author} {\bibfnamefont {C.~L.}\ \bibnamefont
  {Kane}}\ and\ \bibinfo {author} {\bibfnamefont {E.~J.}\ \bibnamefont
  {Mele}},\ }\bibfield  {title} {\enquote {\bibinfo {title} {${Z}_{2}$
  topological order and the quantum spin hall effect},}\ }\href {\doibase
  10.1103/PhysRevLett.95.146802} {\bibfield  {journal} {\bibinfo  {journal}
  {Phys. Rev. Lett.}\ }\textbf {\bibinfo {volume} {95}},\ \bibinfo {pages}
  {146802} (\bibinfo {year} {2005}{\natexlab{a}})}\BibitemShut {NoStop}%
\bibitem [{\citenamefont {Kane}\ and\ \citenamefont
  {Mele}(2005{\natexlab{b}})}]{Kane2005_Quantum}%
  \BibitemOpen
  \bibfield  {author} {\bibinfo {author} {\bibfnamefont {C.~L.}\ \bibnamefont
  {Kane}}\ and\ \bibinfo {author} {\bibfnamefont {E.~J.}\ \bibnamefont
  {Mele}},\ }\bibfield  {title} {\enquote {\bibinfo {title} {Quantum spin hall
  effect in graphene},}\ }\href {\doibase 10.1103/PhysRevLett.95.226801}
  {\bibfield  {journal} {\bibinfo  {journal} {Phys. Rev. Lett.}\ }\textbf
  {\bibinfo {volume} {95}},\ \bibinfo {pages} {226801} (\bibinfo {year}
  {2005}{\natexlab{b}})}\BibitemShut {NoStop}%
\bibitem [{\citenamefont {Bernevig}\ \emph {et~al.}(2006)\citenamefont
  {Bernevig}, \citenamefont {Hughes},\ and\ \citenamefont
  {Zhang}}]{Bernevig_2006}%
  \BibitemOpen
  \bibfield  {author} {\bibinfo {author} {\bibfnamefont {B.}~\bibnamefont
  {Bernevig}}, \bibinfo {author} {\bibfnamefont {T.~L.}\ \bibnamefont
  {Hughes}}, \ and\ \bibinfo {author} {\bibfnamefont {S.-C.}\ \bibnamefont
  {Zhang}},\ }\bibfield  {title} {\enquote {\bibinfo {title} {Quantum spin hall
  insulator state in hgte quantum wells},}\ }\href {http://DOI:
  10.1126/science.1133734} {\bibfield  {journal} {\bibinfo  {journal}
  {Science}\ }\textbf {\bibinfo {volume} {314}},\ \bibinfo {pages} {1757--1761}
  (\bibinfo {year} {2006})}\BibitemShut {NoStop}%
\bibitem [{\citenamefont {Hasan}\ and\ \citenamefont
  {Kane}(2010)}]{Hasan2010_Colloquium}%
  \BibitemOpen
  \bibfield  {author} {\bibinfo {author} {\bibfnamefont {M.~Z.}\ \bibnamefont
  {Hasan}}\ and\ \bibinfo {author} {\bibfnamefont {C.~L.}\ \bibnamefont
  {Kane}},\ }\bibfield  {title} {\enquote {\bibinfo {title} {Colloqium:
  Topological insulators},}\ }\href {\doibase 10.1103/RevModPhys.82.3045}
  {\bibfield  {journal} {\bibinfo  {journal} {Rev. Mod. Phys.}\ }\textbf
  {\bibinfo {volume} {82}},\ \bibinfo {pages} {3045--3067} (\bibinfo {year}
  {2010})}\BibitemShut {NoStop}%
\bibitem [{\citenamefont {Beenakker}(2013)}]{Beenakker2013}%
  \BibitemOpen
  \bibfield  {author} {\bibinfo {author} {\bibfnamefont {C.W.J.}\ \bibnamefont
  {Beenakker}},\ }\bibfield  {title} {\enquote {\bibinfo {title} {Search for
  majorana fermions in superconductors},}\ }\href {\doibase
  10.1146/annurev-conmatphys-030212-184337} {\bibfield  {journal} {\bibinfo
  {journal} {Annual Review of Condensed Matter Physics}\ }\textbf {\bibinfo
  {volume} {4}},\ \bibinfo {pages} {113--136} (\bibinfo {year} {2013})},\
  \Eprint
  {http://arxiv.org/abs/https://doi.org/10.1146/annurev-conmatphys-030212-184337}
  {https://doi.org/10.1146/annurev-conmatphys-030212-184337} \BibitemShut
  {NoStop}%
\bibitem [{\citenamefont {Alicea}(2012)}]{Alicea2012}%
  \BibitemOpen
  \bibfield  {author} {\bibinfo {author} {\bibfnamefont {Jason}\ \bibnamefont
  {Alicea}},\ }\bibfield  {title} {\enquote {\bibinfo {title} {New directions
  in the pursuit of majorana fermions in solid state systems},}\ }\href
  {http://stacks.iop.org/0034-4885/75/i=7/a=076501} {\bibfield  {journal}
  {\bibinfo  {journal} {Reports on Progress in Physics}\ }\textbf {\bibinfo
  {volume} {75}},\ \bibinfo {pages} {076501} (\bibinfo {year}
  {2012})}\BibitemShut {NoStop}%
\bibitem [{\citenamefont {Mourik}\ \emph {et~al.}(2012)\citenamefont {Mourik},
  \citenamefont {Zuo}, \citenamefont {Frolov}, \citenamefont {Plissard},
  \citenamefont {Bakkers},\ and\ \citenamefont {Kouwenhoven}}]{Mourik2012}%
  \BibitemOpen
  \bibfield  {author} {\bibinfo {author} {\bibfnamefont {V.}~\bibnamefont
  {Mourik}}, \bibinfo {author} {\bibfnamefont {K.}~\bibnamefont {Zuo}},
  \bibinfo {author} {\bibfnamefont {S.~M.}\ \bibnamefont {Frolov}}, \bibinfo
  {author} {\bibfnamefont {S.~R.}\ \bibnamefont {Plissard}}, \bibinfo {author}
  {\bibfnamefont {E.~P. A.~M.}\ \bibnamefont {Bakkers}}, \ and\ \bibinfo
  {author} {\bibfnamefont {L.~P.}\ \bibnamefont {Kouwenhoven}},\ }\bibfield
  {title} {\enquote {\bibinfo {title} {Signatures of majorana fermions in
  hybrid superconductor-semiconductor nanowire devices},}\ }\href {\doibase
  10.1126/science.1222360} {\bibfield  {journal} {\bibinfo  {journal}
  {Science}\ }\textbf {\bibinfo {volume} {336}},\ \bibinfo {pages} {1003--1007}
  (\bibinfo {year} {2012})},\ \Eprint
  {http://arxiv.org/abs/http://science.sciencemag.org/content/336/6084/1003.full.pdf}
  {http://science.sciencemag.org/content/336/6084/1003.full.pdf} \BibitemShut
  {NoStop}%
\bibitem [{\citenamefont {Fu}\ and\ \citenamefont {Kane}(2008)}]{Kane2008}%
  \BibitemOpen
  \bibfield  {author} {\bibinfo {author} {\bibfnamefont {Liang}\ \bibnamefont
  {Fu}}\ and\ \bibinfo {author} {\bibfnamefont {C.~L.}\ \bibnamefont {Kane}},\
  }\bibfield  {title} {\enquote {\bibinfo {title} {Superconducting proximity
  effect and majorana fermions at the surface of a topological insulator},}\
  }\href {\doibase 10.1103/PhysRevLett.100.096407} {\bibfield  {journal}
  {\bibinfo  {journal} {Phys. Rev. Lett.}\ }\textbf {\bibinfo {volume} {100}},\
  \bibinfo {pages} {096407} (\bibinfo {year} {2008})}\BibitemShut {NoStop}%
\bibitem [{\citenamefont {Nayak}\ \emph {et~al.}(2008)\citenamefont {Nayak},
  \citenamefont {Simon}, \citenamefont {Stern}, \citenamefont {Freedman},\ and\
  \citenamefont {Das~Sarma}}]{Nayak2008}%
  \BibitemOpen
  \bibfield  {author} {\bibinfo {author} {\bibfnamefont {Chetan}\ \bibnamefont
  {Nayak}}, \bibinfo {author} {\bibfnamefont {Steven~H.}\ \bibnamefont
  {Simon}}, \bibinfo {author} {\bibfnamefont {Ady}\ \bibnamefont {Stern}},
  \bibinfo {author} {\bibfnamefont {Michael}\ \bibnamefont {Freedman}}, \ and\
  \bibinfo {author} {\bibfnamefont {Sankar}\ \bibnamefont {Das~Sarma}},\
  }\bibfield  {title} {\enquote {\bibinfo {title} {Non-abelian anyons and
  topological quantum computation},}\ }\href {\doibase
  10.1103/RevModPhys.80.1083} {\bibfield  {journal} {\bibinfo  {journal} {Rev.
  Mod. Phys.}\ }\textbf {\bibinfo {volume} {80}},\ \bibinfo {pages}
  {1083--1159} (\bibinfo {year} {2008})}\BibitemShut {NoStop}%
\bibitem [{\citenamefont {Das~Sarma}\ \emph {et~al.}(2005)\citenamefont
  {Das~Sarma}, \citenamefont {Freedman},\ and\ \citenamefont
  {Nayak}}]{DasSarma2005}%
  \BibitemOpen
  \bibfield  {author} {\bibinfo {author} {\bibfnamefont {Sankar}\ \bibnamefont
  {Das~Sarma}}, \bibinfo {author} {\bibfnamefont {Michael}\ \bibnamefont
  {Freedman}}, \ and\ \bibinfo {author} {\bibfnamefont {Chetan}\ \bibnamefont
  {Nayak}},\ }\bibfield  {title} {\enquote {\bibinfo {title} {Topologically
  protected qubits from a possible non-abelian fractional quantum hall
  state},}\ }\href {\doibase 10.1103/PhysRevLett.94.166802} {\bibfield
  {journal} {\bibinfo  {journal} {Phys. Rev. Lett.}\ }\textbf {\bibinfo
  {volume} {94}},\ \bibinfo {pages} {166802} (\bibinfo {year}
  {2005})}\BibitemShut {NoStop}%
\bibitem [{\citenamefont {Jotzu}\ \emph {et~al.}(2014)\citenamefont {Jotzu},
  \citenamefont {Messer}, \citenamefont {Desbuquois}, \citenamefont {Lebrat},
  \citenamefont {Uehlinger}, \citenamefont {Greif},\ and\ \citenamefont
  {Esslinger}}]{Jotzu2014_Experimental}%
  \BibitemOpen
  \bibfield  {author} {\bibinfo {author} {\bibfnamefont {Gregor}\ \bibnamefont
  {Jotzu}}, \bibinfo {author} {\bibfnamefont {Michael}\ \bibnamefont {Messer}},
  \bibinfo {author} {\bibfnamefont {R{\'e}mi}\ \bibnamefont {Desbuquois}},
  \bibinfo {author} {\bibfnamefont {Martin}\ \bibnamefont {Lebrat}}, \bibinfo
  {author} {\bibfnamefont {Thomas}\ \bibnamefont {Uehlinger}}, \bibinfo
  {author} {\bibfnamefont {Daniel}\ \bibnamefont {Greif}}, \ and\ \bibinfo
  {author} {\bibfnamefont {Tilman}\ \bibnamefont {Esslinger}},\ }\bibfield
  {title} {\enquote {\bibinfo {title} {Experimental realization of the
  topological haldane model with ultracold fermions},}\ }\href
  {http://dx.doi.org/10.1038/nature13915} {\bibfield  {journal} {\bibinfo
  {journal} {Nature}\ }\textbf {\bibinfo {volume} {515}},\ \bibinfo {pages}
  {237 EP --} (\bibinfo {year} {2014})}\BibitemShut {NoStop}%
\bibitem [{\citenamefont {Lanneb\`ere}\ and\ \citenamefont
  {Silveirinha}(2018)}]{Lannebere2018_Link}%
  \BibitemOpen
  \bibfield  {author} {\bibinfo {author} {\bibfnamefont {Sylvain}\ \bibnamefont
  {Lanneb\`ere}}\ and\ \bibinfo {author} {\bibfnamefont {M\'ario~G.}\
  \bibnamefont {Silveirinha}},\ }\bibfield  {title} {\enquote {\bibinfo {title}
  {Link between the photonic and electronic topological phases in artificial
  graphene},}\ }\href {\doibase 10.1103/PhysRevB.97.165128} {\bibfield
  {journal} {\bibinfo  {journal} {Phys. Rev. B}\ }\textbf {\bibinfo {volume}
  {97}},\ \bibinfo {pages} {165128} (\bibinfo {year} {2018})}\BibitemShut
  {NoStop}%
\bibitem [{\citenamefont {Liu}\ \emph {et~al.}(2017{\natexlab{a}})\citenamefont
  {Liu}, \citenamefont {Xu}, \citenamefont {Zhang},\ and\ \citenamefont
  {Duan}}]{Liu2017_Model}%
  \BibitemOpen
  \bibfield  {author} {\bibinfo {author} {\bibfnamefont {Yizhou}\ \bibnamefont
  {Liu}}, \bibinfo {author} {\bibfnamefont {Yong}\ \bibnamefont {Xu}}, \bibinfo
  {author} {\bibfnamefont {Shou-Cheng}\ \bibnamefont {Zhang}}, \ and\ \bibinfo
  {author} {\bibfnamefont {Wenhui}\ \bibnamefont {Duan}},\ }\bibfield  {title}
  {\enquote {\bibinfo {title} {Model for topological phononics and phonon
  diode},}\ }\href {\doibase 10.1103/PhysRevB.96.064106} {\bibfield  {journal}
  {\bibinfo  {journal} {Phys. Rev. B}\ }\textbf {\bibinfo {volume} {96}},\
  \bibinfo {pages} {064106} (\bibinfo {year} {2017}{\natexlab{a}})}\BibitemShut
  {NoStop}%
\bibitem [{\citenamefont {Owerre}(2016{\natexlab{a}})}]{Owerre2016_First}%
  \BibitemOpen
  \bibfield  {author} {\bibinfo {author} {\bibfnamefont {S~A}\ \bibnamefont
  {Owerre}},\ }\bibfield  {title} {\enquote {\bibinfo {title} {A first
  theoretical realization of honeycomb topological magnon insulator},}\ }\href
  {http://stacks.iop.org/0953-8984/28/i=38/a=386001} {\bibfield  {journal}
  {\bibinfo  {journal} {Journal of Physics: Condensed Matter}\ }\textbf
  {\bibinfo {volume} {28}},\ \bibinfo {pages} {386001} (\bibinfo {year}
  {2016}{\natexlab{a}})}\BibitemShut {NoStop}%
\bibitem [{\citenamefont
  {Owerre}(2016{\natexlab{b}})}]{Owerre2016_Topological}%
  \BibitemOpen
  \bibfield  {author} {\bibinfo {author} {\bibfnamefont {S.~A.}\ \bibnamefont
  {Owerre}},\ }\bibfield  {title} {\enquote {\bibinfo {title} {Topological
  honeycomb magnon hall effect: A calculation of thermal hall conductivity of
  magnetic spin excitations},}\ }\href {\doibase 10.1063/1.4959815} {\bibfield
  {journal} {\bibinfo  {journal} {Journal of Applied Physics}\ }\textbf
  {\bibinfo {volume} {120}},\ \bibinfo {pages} {043903} (\bibinfo {year}
  {2016}{\natexlab{b}})},\ \Eprint
  {http://arxiv.org/abs/https://doi.org/10.1063/1.4959815}
  {https://doi.org/10.1063/1.4959815} \BibitemShut {NoStop}%
\bibitem [{\citenamefont {Kim}\ \emph {et~al.}(2016)\citenamefont {Kim},
  \citenamefont {Ochoa}, \citenamefont {Zarzuela},\ and\ \citenamefont
  {Tserkovnyak}}]{Kim2016_Realization}%
  \BibitemOpen
  \bibfield  {author} {\bibinfo {author} {\bibfnamefont {Se~Kwon}\ \bibnamefont
  {Kim}}, \bibinfo {author} {\bibfnamefont {H\'ector}\ \bibnamefont {Ochoa}},
  \bibinfo {author} {\bibfnamefont {Ricardo}\ \bibnamefont {Zarzuela}}, \ and\
  \bibinfo {author} {\bibfnamefont {Yaroslav}\ \bibnamefont {Tserkovnyak}},\
  }\bibfield  {title} {\enquote {\bibinfo {title} {Realization of the
  haldane-kane-mele model in a system of localized spins},}\ }\href {\doibase
  10.1103/PhysRevLett.117.227201} {\bibfield  {journal} {\bibinfo  {journal}
  {Phys. Rev. Lett.}\ }\textbf {\bibinfo {volume} {117}},\ \bibinfo {pages}
  {227201} (\bibinfo {year} {2016})}\BibitemShut {NoStop}%
\bibitem [{\citenamefont {Owerre}\ and\ \citenamefont
  {Nsofini}(2017)}]{Owerre2017_Squeezed}%
  \BibitemOpen
  \bibfield  {author} {\bibinfo {author} {\bibfnamefont {S~A}\ \bibnamefont
  {Owerre}}\ and\ \bibinfo {author} {\bibfnamefont {J}~\bibnamefont
  {Nsofini}},\ }\bibfield  {title} {\enquote {\bibinfo {title} {Squeezed dirac
  and topological magnons in a bosonic honeycomb optical lattice},}\ }\href
  {http://stacks.iop.org/0953-8984/29/i=45/a=455802} {\bibfield  {journal}
  {\bibinfo  {journal} {Journal of Physics: Condensed Matter}\ }\textbf
  {\bibinfo {volume} {29}},\ \bibinfo {pages} {455802} (\bibinfo {year}
  {2017})}\BibitemShut {NoStop}%
\bibitem [{\citenamefont {Kim}\ and\ \citenamefont
  {Kee}(2017)}]{Kim2017_Realizing}%
  \BibitemOpen
  \bibfield  {author} {\bibinfo {author} {\bibfnamefont {Heung-Sik}\
  \bibnamefont {Kim}}\ and\ \bibinfo {author} {\bibfnamefont {Hae-Young}\
  \bibnamefont {Kee}},\ }\bibfield  {title} {\enquote {\bibinfo {title}
  {Realizing haldane model in fe-based honeycomb ferromagnetic insulators},}\
  }\href {\doibase 10.1038/s41535-017-0021-z} {\bibfield  {journal} {\bibinfo
  {journal} {npj Quantum Materials}\ }\textbf {\bibinfo {volume} {2}},\
  \bibinfo {pages} {20} (\bibinfo {year} {2017})}\BibitemShut {NoStop}%
\bibitem [{\citenamefont
  {Dzyaloshinsky}(1958)}]{Dzyaloshinskii1958_Thermodynamic}%
  \BibitemOpen
  \bibfield  {author} {\bibinfo {author} {\bibfnamefont {I.}~\bibnamefont
  {Dzyaloshinsky}},\ }\bibfield  {title} {\enquote {\bibinfo {title} {A
  thermodynamic theory of “weak” ferromagnetism of antiferromagnetics},}\
  }\href {\doibase https://doi.org/10.1016/0022-3697(58)90076-3} {\bibfield
  {journal} {\bibinfo  {journal} {Journal of Physics and Chemistry of Solids}\
  }\textbf {\bibinfo {volume} {4}},\ \bibinfo {pages} {241 -- 255} (\bibinfo
  {year} {1958})}\BibitemShut {NoStop}%
\bibitem [{\citenamefont {Moriya}(1960)}]{Moriya1960_Anisotropic}%
  \BibitemOpen
  \bibfield  {author} {\bibinfo {author} {\bibfnamefont {T\^oru}\ \bibnamefont
  {Moriya}},\ }\bibfield  {title} {\enquote {\bibinfo {title} {Anisotropic
  superexchange interaction and weak ferromagnetism},}\ }\href {\doibase
  10.1103/PhysRev.120.91} {\bibfield  {journal} {\bibinfo  {journal} {Phys.
  Rev.}\ }\textbf {\bibinfo {volume} {120}},\ \bibinfo {pages} {91--98}
  (\bibinfo {year} {1960})}\BibitemShut {NoStop}%
\bibitem [{\citenamefont {R\"uckriegel}\ \emph {et~al.}(2018)\citenamefont
  {R\"uckriegel}, \citenamefont {Brataas},\ and\ \citenamefont
  {Duine}}]{Ruckriegel2017_Bulk}%
  \BibitemOpen
  \bibfield  {author} {\bibinfo {author} {\bibfnamefont {Andreas}\ \bibnamefont
  {R\"uckriegel}}, \bibinfo {author} {\bibfnamefont {Arne}\ \bibnamefont
  {Brataas}}, \ and\ \bibinfo {author} {\bibfnamefont {Rembert~A.}\
  \bibnamefont {Duine}},\ }\bibfield  {title} {\enquote {\bibinfo {title} {Bulk
  and edge spin transport in topological magnon insulators},}\ }\href {\doibase
  10.1103/PhysRevB.97.081106} {\bibfield  {journal} {\bibinfo  {journal} {Phys.
  Rev. B}\ }\textbf {\bibinfo {volume} {97}},\ \bibinfo {pages} {081106}
  (\bibinfo {year} {2018})}\BibitemShut {NoStop}%
\bibitem [{\citenamefont {Kruglyak}\ \emph {et~al.}(2010)\citenamefont
  {Kruglyak}, \citenamefont {Demokritov},\ and\ \citenamefont
  {Grundler}}]{Kruglyak2010}%
  \BibitemOpen
  \bibfield  {author} {\bibinfo {author} {\bibfnamefont {V~V}\ \bibnamefont
  {Kruglyak}}, \bibinfo {author} {\bibfnamefont {S~O}\ \bibnamefont
  {Demokritov}}, \ and\ \bibinfo {author} {\bibfnamefont {D}~\bibnamefont
  {Grundler}},\ }\bibfield  {title} {\enquote {\bibinfo {title} {Magnonics},}\
  }\href {http://stacks.iop.org/0022-3727/43/i=26/a=264001} {\bibfield
  {journal} {\bibinfo  {journal} {Journal of Physics D: Applied Physics}\
  }\textbf {\bibinfo {volume} {43}},\ \bibinfo {pages} {264001} (\bibinfo
  {year} {2010})}\BibitemShut {NoStop}%
\bibitem [{\citenamefont {Chumak}\ \emph {et~al.}(2015)\citenamefont {Chumak},
  \citenamefont {Vasyuchka}, \citenamefont {Serga},\ and\ \citenamefont
  {Hillebrands}}]{Chumak2015}%
  \BibitemOpen
  \bibfield  {author} {\bibinfo {author} {\bibfnamefont {A.~V.}\ \bibnamefont
  {Chumak}}, \bibinfo {author} {\bibfnamefont {V.~I.}\ \bibnamefont
  {Vasyuchka}}, \bibinfo {author} {\bibfnamefont {A.~A.}\ \bibnamefont
  {Serga}}, \ and\ \bibinfo {author} {\bibfnamefont {B.}~\bibnamefont
  {Hillebrands}},\ }\bibfield  {title} {\enquote {\bibinfo {title} {Magnon
  spintronics},}\ }\href {http://dx.doi.org/10.1038/nphys3347} {\bibfield
  {journal} {\bibinfo  {journal} {Nat Phys}\ }\textbf {\bibinfo {volume}
  {11}},\ \bibinfo {pages} {453} (\bibinfo {year} {2015})}\BibitemShut
  {NoStop}%
\bibitem [{\citenamefont {Bauer}\ \emph {et~al.}(2012)\citenamefont {Bauer},
  \citenamefont {Saitoh},\ and\ \citenamefont {van Wees}}]{Bauer2012}%
  \BibitemOpen
  \bibfield  {author} {\bibinfo {author} {\bibfnamefont {Gerrit E.~W.}\
  \bibnamefont {Bauer}}, \bibinfo {author} {\bibfnamefont {Eiji}\ \bibnamefont
  {Saitoh}}, \ and\ \bibinfo {author} {\bibfnamefont {Bart~J.}\ \bibnamefont
  {van Wees}},\ }\bibfield  {title} {\enquote {\bibinfo {title} {Spin
  caloritronics},}\ }\href {\doibase http://dx.doi.org/10.1038/nmat3301}
  {\bibfield  {journal} {\bibinfo  {journal} {Nat Mater}\ }\textbf {\bibinfo
  {volume} {11}},\ \bibinfo {pages} {391} (\bibinfo {year} {2012})}\BibitemShut
  {NoStop}%
\bibitem [{\citenamefont {Uchida}\ \emph {et~al.}(2010)\citenamefont {Uchida},
  \citenamefont {Xiao}, \citenamefont {Adachi}, \citenamefont {Ohe},
  \citenamefont {Takahashi}, \citenamefont {Ieda}, \citenamefont {Ota},
  \citenamefont {Kajiwara}, \citenamefont {Umezawa}, \citenamefont {Kawai},
  \citenamefont {Bauer}, \citenamefont {Maekawa},\ and\ \citenamefont
  {Saitoh}}]{Uchida2010}%
  \BibitemOpen
  \bibfield  {author} {\bibinfo {author} {\bibfnamefont {K.}~\bibnamefont
  {Uchida}}, \bibinfo {author} {\bibfnamefont {J.}~\bibnamefont {Xiao}},
  \bibinfo {author} {\bibfnamefont {H.}~\bibnamefont {Adachi}}, \bibinfo
  {author} {\bibfnamefont {J.}~\bibnamefont {Ohe}}, \bibinfo {author}
  {\bibfnamefont {S.}~\bibnamefont {Takahashi}}, \bibinfo {author}
  {\bibfnamefont {J.}~\bibnamefont {Ieda}}, \bibinfo {author} {\bibfnamefont
  {T.}~\bibnamefont {Ota}}, \bibinfo {author} {\bibfnamefont {Y.}~\bibnamefont
  {Kajiwara}}, \bibinfo {author} {\bibfnamefont {H.}~\bibnamefont {Umezawa}},
  \bibinfo {author} {\bibfnamefont {H.}~\bibnamefont {Kawai}}, \bibinfo
  {author} {\bibfnamefont {G.~E.~W.}\ \bibnamefont {Bauer}}, \bibinfo {author}
  {\bibfnamefont {S.}~\bibnamefont {Maekawa}}, \ and\ \bibinfo {author}
  {\bibfnamefont {E.}~\bibnamefont {Saitoh}},\ }\bibfield  {title} {\enquote
  {\bibinfo {title} {Spin seebeck insulator},}\ }\href {\doibase
  10.1038/nmat2856} {\bibfield  {journal} {\bibinfo  {journal} {Nat Mater}\
  }\textbf {\bibinfo {volume} {9}},\ \bibinfo {pages} {894--897} (\bibinfo
  {year} {2010})}\BibitemShut {NoStop}%
\bibitem [{\citenamefont {Akhiezer}\ \emph {et~al.}(1968)\citenamefont
  {Akhiezer}, \citenamefont {Bar'iakhtar},\ and\ \citenamefont
  {Peletminski}}]{Akhiezer1968}%
  \BibitemOpen
  \bibfield  {author} {\bibinfo {author} {\bibfnamefont {A.I.}\ \bibnamefont
  {Akhiezer}}, \bibinfo {author} {\bibfnamefont {V.G.}\ \bibnamefont
  {Bar'iakhtar}}, \ and\ \bibinfo {author} {\bibfnamefont {S.V.}\ \bibnamefont
  {Peletminski}},\ }\href {http://books.google.nl/books?id=GpA6AAAAMAAJ} {\emph
  {\bibinfo {title} {Spin waves}}}\ (\bibinfo  {publisher} {North-Holland
  Publishing Company, Amsterdam},\ \bibinfo {year} {1968})\BibitemShut
  {NoStop}%
\bibitem [{\citenamefont {Chumak}\ \emph {et~al.}(2014)\citenamefont {Chumak},
  \citenamefont {Serga},\ and\ \citenamefont {Hillebrands}}]{Chumak2014}%
  \BibitemOpen
  \bibfield  {author} {\bibinfo {author} {\bibfnamefont {Andrii~V.}\
  \bibnamefont {Chumak}}, \bibinfo {author} {\bibfnamefont {Alexander~A.}\
  \bibnamefont {Serga}}, \ and\ \bibinfo {author} {\bibfnamefont {Burkard}\
  \bibnamefont {Hillebrands}},\ }\bibfield  {title} {\enquote {\bibinfo {title}
  {Magnon transistor for all-magnon data processing},}\ }\href {\doibase
  10.1038/ncomms5700} {\bibfield  {journal} {\bibinfo  {journal} {Nature
  Communications}\ }\textbf {\bibinfo {volume} {5}},\ \bibinfo {pages} {4700}
  (\bibinfo {year} {2014})}\BibitemShut {NoStop}%
\bibitem [{\citenamefont {Ganzhorn}\ \emph {et~al.}(2016)\citenamefont
  {Ganzhorn}, \citenamefont {Klingler}, \citenamefont {Wimmer}, \citenamefont
  {Geprägs}, \citenamefont {Gross}, \citenamefont {Huebl},\ and\ \citenamefont
  {Goennenwein}}]{Ganzhorn2016}%
  \BibitemOpen
  \bibfield  {author} {\bibinfo {author} {\bibfnamefont {Kathrin}\ \bibnamefont
  {Ganzhorn}}, \bibinfo {author} {\bibfnamefont {Stefan}\ \bibnamefont
  {Klingler}}, \bibinfo {author} {\bibfnamefont {Tobias}\ \bibnamefont
  {Wimmer}}, \bibinfo {author} {\bibfnamefont {Stephan}\ \bibnamefont
  {Geprägs}}, \bibinfo {author} {\bibfnamefont {Rudolf}\ \bibnamefont
  {Gross}}, \bibinfo {author} {\bibfnamefont {Hans}\ \bibnamefont {Huebl}}, \
  and\ \bibinfo {author} {\bibfnamefont {Sebastian T.~B.}\ \bibnamefont
  {Goennenwein}},\ }\bibfield  {title} {\enquote {\bibinfo {title}
  {Magnon-based logic in a multi-terminal yig/pt nanostructure},}\ }\href
  {\doibase 10.1063/1.4958893} {\bibfield  {journal} {\bibinfo  {journal}
  {Applied Physics Letters}\ }\textbf {\bibinfo {volume} {109}},\ \bibinfo
  {pages} {022405} (\bibinfo {year} {2016})},\ \Eprint
  {http://arxiv.org/abs/https://doi.org/10.1063/1.4958893}
  {https://doi.org/10.1063/1.4958893} \BibitemShut {NoStop}%
\bibitem [{\citenamefont {An}\ \emph {et~al.}(2013)\citenamefont {An},
  \citenamefont {Vasyuchka}, \citenamefont {Uchida}, \citenamefont {Chumak},
  \citenamefont {Yamaguchi}, \citenamefont {Harii}, \citenamefont {Ohe},
  \citenamefont {Jungfleisch}, \citenamefont {Kajiwara}, \citenamefont
  {Adachi}, \citenamefont {Hillebrands}, \citenamefont {Maekawa},\ and\
  \citenamefont {Saitoh}}]{An2013}%
  \BibitemOpen
  \bibfield  {author} {\bibinfo {author} {\bibfnamefont {T.}~\bibnamefont
  {An}}, \bibinfo {author} {\bibfnamefont {V.~I.}\ \bibnamefont {Vasyuchka}},
  \bibinfo {author} {\bibfnamefont {K.}~\bibnamefont {Uchida}}, \bibinfo
  {author} {\bibfnamefont {A.~V.}\ \bibnamefont {Chumak}}, \bibinfo {author}
  {\bibfnamefont {K.}~\bibnamefont {Yamaguchi}}, \bibinfo {author}
  {\bibfnamefont {K.}~\bibnamefont {Harii}}, \bibinfo {author} {\bibfnamefont
  {J.}~\bibnamefont {Ohe}}, \bibinfo {author} {\bibfnamefont {M.~B.}\
  \bibnamefont {Jungfleisch}}, \bibinfo {author} {\bibfnamefont
  {Y.}~\bibnamefont {Kajiwara}}, \bibinfo {author} {\bibfnamefont
  {H.}~\bibnamefont {Adachi}}, \bibinfo {author} {\bibfnamefont
  {B.}~\bibnamefont {Hillebrands}}, \bibinfo {author} {\bibfnamefont
  {S.}~\bibnamefont {Maekawa}}, \ and\ \bibinfo {author} {\bibfnamefont
  {E.}~\bibnamefont {Saitoh}},\ }\bibfield  {title} {\enquote {\bibinfo {title}
  {Unidirectional spin-wave heat conveyer},}\ }\href {\doibase
  10.1038/nmat3628} {\bibfield  {journal} {\bibinfo  {journal} {Nature
  Materials}\ }\textbf {\bibinfo {volume} {12}},\ \bibinfo {pages} {549}
  (\bibinfo {year} {2013})}\BibitemShut {NoStop}%
\bibitem [{\citenamefont {Saitoh}\ \emph {et~al.}(2006)\citenamefont {Saitoh},
  \citenamefont {Ueda}, \citenamefont {Miyajima},\ and\ \citenamefont
  {Tatara}}]{Saitoh2006}%
  \BibitemOpen
  \bibfield  {author} {\bibinfo {author} {\bibfnamefont {E.}~\bibnamefont
  {Saitoh}}, \bibinfo {author} {\bibfnamefont {M.}~\bibnamefont {Ueda}},
  \bibinfo {author} {\bibfnamefont {H.}~\bibnamefont {Miyajima}}, \ and\
  \bibinfo {author} {\bibfnamefont {G.}~\bibnamefont {Tatara}},\ }\bibfield
  {title} {\enquote {\bibinfo {title} {Conversion of spin current into charge
  current at room temperature: Inverse spin-hall effect},}\ }\href {\doibase
  http://dx.doi.org/10.1063/1.2199473} {\bibfield  {journal} {\bibinfo
  {journal} {Applied Physics Letters}\ }\textbf {\bibinfo {volume} {88}},\
  \bibinfo {pages} {182509} (\bibinfo {year} {2006})}\BibitemShut {NoStop}%
\bibitem [{\citenamefont {Sonin}(2010)}]{Sonin2010}%
  \BibitemOpen
  \bibfield  {author} {\bibinfo {author} {\bibfnamefont {E.B.}\ \bibnamefont
  {Sonin}},\ }\bibfield  {title} {\enquote {\bibinfo {title} {Spin currents and
  spin superfluidity},}\ }\href {\doibase 10.1080/00018731003739943} {\bibfield
   {journal} {\bibinfo  {journal} {Advances in Physics}\ }\textbf {\bibinfo
  {volume} {59}},\ \bibinfo {pages} {181--255} (\bibinfo {year} {2010})},\
  \Eprint {http://arxiv.org/abs/http://dx.doi.org/10.1080/00018731003739943}
  {http://dx.doi.org/10.1080/00018731003739943} \BibitemShut {NoStop}%
\bibitem [{\citenamefont {Takei}\ \emph {et~al.}(2014)\citenamefont {Takei},
  \citenamefont {Halperin}, \citenamefont {Yacoby},\ and\ \citenamefont
  {Tserkovnyak}}]{Takei2014}%
  \BibitemOpen
  \bibfield  {author} {\bibinfo {author} {\bibfnamefont {So}~\bibnamefont
  {Takei}}, \bibinfo {author} {\bibfnamefont {Bertrand~I.}\ \bibnamefont
  {Halperin}}, \bibinfo {author} {\bibfnamefont {Amir}\ \bibnamefont {Yacoby}},
  \ and\ \bibinfo {author} {\bibfnamefont {Yaroslav}\ \bibnamefont
  {Tserkovnyak}},\ }\bibfield  {title} {\enquote {\bibinfo {title} {Superfluid
  spin transport through antiferromagnetic insulators},}\ }\href {\doibase
  10.1103/PhysRevB.90.094408} {\bibfield  {journal} {\bibinfo  {journal} {Phys.
  Rev. B}\ }\textbf {\bibinfo {volume} {90}},\ \bibinfo {pages} {094408}
  (\bibinfo {year} {2014})}\BibitemShut {NoStop}%
\bibitem [{\citenamefont {Kamra}\ and\ \citenamefont
  {Belzig}(2016)}]{Kamra2016A}%
  \BibitemOpen
  \bibfield  {author} {\bibinfo {author} {\bibfnamefont {Akashdeep}\
  \bibnamefont {Kamra}}\ and\ \bibinfo {author} {\bibfnamefont {Wolfgang}\
  \bibnamefont {Belzig}},\ }\bibfield  {title} {\enquote {\bibinfo {title}
  {Super-poissonian shot noise of squeezed-magnon mediated spin transport},}\
  }\href {\doibase 10.1103/PhysRevLett.116.146601} {\bibfield  {journal}
  {\bibinfo  {journal} {Phys. Rev. Lett.}\ }\textbf {\bibinfo {volume} {116}},\
  \bibinfo {pages} {146601} (\bibinfo {year} {2016})}\BibitemShut {NoStop}%
\bibitem [{\citenamefont {Kamra}\ \emph {et~al.}(2017)\citenamefont {Kamra},
  \citenamefont {Agrawal},\ and\ \citenamefont {Belzig}}]{Kamra2017}%
  \BibitemOpen
  \bibfield  {author} {\bibinfo {author} {\bibfnamefont {Akashdeep}\
  \bibnamefont {Kamra}}, \bibinfo {author} {\bibfnamefont {Utkarsh}\
  \bibnamefont {Agrawal}}, \ and\ \bibinfo {author} {\bibfnamefont {Wolfgang}\
  \bibnamefont {Belzig}},\ }\bibfield  {title} {\enquote {\bibinfo {title}
  {Noninteger-spin magnonic excitations in untextured magnets},}\ }\href
  {\doibase 10.1103/PhysRevB.96.020411} {\bibfield  {journal} {\bibinfo
  {journal} {Phys. Rev. B}\ }\textbf {\bibinfo {volume} {96}},\ \bibinfo
  {pages} {020411} (\bibinfo {year} {2017})}\BibitemShut {NoStop}%
\bibitem [{\citenamefont {Kittel}(1949)}]{Kittel1949}%
  \BibitemOpen
  \bibfield  {author} {\bibinfo {author} {\bibfnamefont {Charles}\ \bibnamefont
  {Kittel}},\ }\bibfield  {title} {\enquote {\bibinfo {title} {Physical theory
  of ferromagnetic domains},}\ }\href {\doibase 10.1103/RevModPhys.21.541}
  {\bibfield  {journal} {\bibinfo  {journal} {Rev. Mod. Phys.}\ }\textbf
  {\bibinfo {volume} {21}},\ \bibinfo {pages} {541--583} (\bibinfo {year}
  {1949})}\BibitemShut {NoStop}%
\bibitem [{\citenamefont {Kittel}(1958)}]{Kittel1958}%
  \BibitemOpen
  \bibfield  {author} {\bibinfo {author} {\bibfnamefont {C.}~\bibnamefont
  {Kittel}},\ }\bibfield  {title} {\enquote {\bibinfo {title} {Interaction of
  spin waves and ultrasonic waves in ferromagnetic crystals},}\ }\href
  {\doibase 10.1103/PhysRev.110.836} {\bibfield  {journal} {\bibinfo  {journal}
  {Phys. Rev.}\ }\textbf {\bibinfo {volume} {110}},\ \bibinfo {pages}
  {836--841} (\bibinfo {year} {1958})}\BibitemShut {NoStop}%
\bibitem [{\citenamefont {Uchida}\ \emph {et~al.}(2011)\citenamefont {Uchida},
  \citenamefont {Adachi}, \citenamefont {An}, \citenamefont {Ota},
  \citenamefont {Toda}, \citenamefont {Hillebrands}, \citenamefont {Maekawa},\
  and\ \citenamefont {Saitoh}}]{Uchida2011}%
  \BibitemOpen
  \bibfield  {author} {\bibinfo {author} {\bibfnamefont {K.}~\bibnamefont
  {Uchida}}, \bibinfo {author} {\bibfnamefont {H.}~\bibnamefont {Adachi}},
  \bibinfo {author} {\bibfnamefont {T.}~\bibnamefont {An}}, \bibinfo {author}
  {\bibfnamefont {T.}~\bibnamefont {Ota}}, \bibinfo {author} {\bibfnamefont
  {M.}~\bibnamefont {Toda}}, \bibinfo {author} {\bibfnamefont {B.}~\bibnamefont
  {Hillebrands}}, \bibinfo {author} {\bibfnamefont {S.}~\bibnamefont
  {Maekawa}}, \ and\ \bibinfo {author} {\bibfnamefont {E.}~\bibnamefont
  {Saitoh}},\ }\bibfield  {title} {\enquote {\bibinfo {title} {Long-range spin
  seebeck effect and acoustic spin pumping},}\ }\href {\doibase
  10.1038/nmat3099} {\bibfield  {journal} {\bibinfo  {journal} {Nat Mater}\
  }\textbf {\bibinfo {volume} {10}},\ \bibinfo {pages} {737} (\bibinfo {year}
  {2011})}\BibitemShut {NoStop}%
\bibitem [{\citenamefont {Kamra}\ \emph {et~al.}(2015)\citenamefont {Kamra},
  \citenamefont {Keshtgar}, \citenamefont {Yan},\ and\ \citenamefont
  {Bauer}}]{Kamra2015_Coherent}%
  \BibitemOpen
  \bibfield  {author} {\bibinfo {author} {\bibfnamefont {Akashdeep}\
  \bibnamefont {Kamra}}, \bibinfo {author} {\bibfnamefont {Hedyeh}\
  \bibnamefont {Keshtgar}}, \bibinfo {author} {\bibfnamefont {Peng}\
  \bibnamefont {Yan}}, \ and\ \bibinfo {author} {\bibfnamefont {Gerrit E.~W.}\
  \bibnamefont {Bauer}},\ }\bibfield  {title} {\enquote {\bibinfo {title}
  {Coherent elastic excitation of spin waves},}\ }\href {\doibase
  10.1103/PhysRevB.91.104409} {\bibfield  {journal} {\bibinfo  {journal} {Phys.
  Rev. B}\ }\textbf {\bibinfo {volume} {91}},\ \bibinfo {pages} {104409}
  (\bibinfo {year} {2015})}\BibitemShut {NoStop}%
\bibitem [{\citenamefont {Kamra}\ and\ \citenamefont
  {Bauer}(2014)}]{Kamra2014_Actuation}%
  \BibitemOpen
  \bibfield  {author} {\bibinfo {author} {\bibfnamefont {Akashdeep}\
  \bibnamefont {Kamra}}\ and\ \bibinfo {author} {\bibfnamefont {Gerrit~E.W.}\
  \bibnamefont {Bauer}},\ }\bibfield  {title} {\enquote {\bibinfo {title}
  {Actuation, propagation, and detection of transverse magnetoelastic waves in
  ferromagnets},}\ }\href {\doibase https://doi.org/10.1016/j.ssc.2013.10.007}
  {\bibfield  {journal} {\bibinfo  {journal} {Solid State Communications}\
  }\textbf {\bibinfo {volume} {198}},\ \bibinfo {pages} {35 -- 39} (\bibinfo
  {year} {2014})},\ \bibinfo {note} {sI: Spin Mechanics}\BibitemShut {NoStop}%
\bibitem [{\citenamefont {Weiler}\ \emph {et~al.}(2012)\citenamefont {Weiler},
  \citenamefont {Huebl}, \citenamefont {Goerg}, \citenamefont {Czeschka},
  \citenamefont {Gross},\ and\ \citenamefont {Goennenwein}}]{Weiler2012}%
  \BibitemOpen
  \bibfield  {author} {\bibinfo {author} {\bibfnamefont {M.}~\bibnamefont
  {Weiler}}, \bibinfo {author} {\bibfnamefont {H.}~\bibnamefont {Huebl}},
  \bibinfo {author} {\bibfnamefont {F.~S.}\ \bibnamefont {Goerg}}, \bibinfo
  {author} {\bibfnamefont {F.~D.}\ \bibnamefont {Czeschka}}, \bibinfo {author}
  {\bibfnamefont {R.}~\bibnamefont {Gross}}, \ and\ \bibinfo {author}
  {\bibfnamefont {S.~T.~B.}\ \bibnamefont {Goennenwein}},\ }\bibfield  {title}
  {\enquote {\bibinfo {title} {Spin pumping with coherent elastic waves},}\
  }\href {\doibase 10.1103/PhysRevLett.108.176601} {\bibfield  {journal}
  {\bibinfo  {journal} {Phys. Rev. Lett.}\ }\textbf {\bibinfo {volume} {108}},\
  \bibinfo {pages} {176601} (\bibinfo {year} {2012})}\BibitemShut {NoStop}%
\bibitem [{\citenamefont {Dreher}\ \emph {et~al.}(2012)\citenamefont {Dreher},
  \citenamefont {Weiler}, \citenamefont {Pernpeintner}, \citenamefont {Huebl},
  \citenamefont {Gross}, \citenamefont {Brandt},\ and\ \citenamefont
  {Goennenwein}}]{Dreher2012}%
  \BibitemOpen
  \bibfield  {author} {\bibinfo {author} {\bibfnamefont {L.}~\bibnamefont
  {Dreher}}, \bibinfo {author} {\bibfnamefont {M.}~\bibnamefont {Weiler}},
  \bibinfo {author} {\bibfnamefont {M.}~\bibnamefont {Pernpeintner}}, \bibinfo
  {author} {\bibfnamefont {H.}~\bibnamefont {Huebl}}, \bibinfo {author}
  {\bibfnamefont {R.}~\bibnamefont {Gross}}, \bibinfo {author} {\bibfnamefont
  {M.~S.}\ \bibnamefont {Brandt}}, \ and\ \bibinfo {author} {\bibfnamefont
  {S.~T.~B.}\ \bibnamefont {Goennenwein}},\ }\bibfield  {title} {\enquote
  {\bibinfo {title} {Surface acoustic wave driven ferromagnetic resonance in
  nickel thin films: Theory and experiment},}\ }\href {\doibase
  10.1103/PhysRevB.86.134415} {\bibfield  {journal} {\bibinfo  {journal} {Phys.
  Rev. B}\ }\textbf {\bibinfo {volume} {86}},\ \bibinfo {pages} {134415}
  (\bibinfo {year} {2012})}\BibitemShut {NoStop}%
\bibitem [{\citenamefont {Flebus}\ \emph {et~al.}(2017)\citenamefont {Flebus},
  \citenamefont {Shen}, \citenamefont {Kikkawa}, \citenamefont {Uchida},
  \citenamefont {Qiu}, \citenamefont {Saitoh}, \citenamefont {Duine},\ and\
  \citenamefont {Bauer}}]{Flebus2017}%
  \BibitemOpen
  \bibfield  {author} {\bibinfo {author} {\bibfnamefont {Benedetta}\
  \bibnamefont {Flebus}}, \bibinfo {author} {\bibfnamefont {Ka}~\bibnamefont
  {Shen}}, \bibinfo {author} {\bibfnamefont {Takashi}\ \bibnamefont {Kikkawa}},
  \bibinfo {author} {\bibfnamefont {Ken-ichi}\ \bibnamefont {Uchida}}, \bibinfo
  {author} {\bibfnamefont {Zhiyong}\ \bibnamefont {Qiu}}, \bibinfo {author}
  {\bibfnamefont {Eiji}\ \bibnamefont {Saitoh}}, \bibinfo {author}
  {\bibfnamefont {Rembert~A.}\ \bibnamefont {Duine}}, \ and\ \bibinfo {author}
  {\bibfnamefont {Gerrit E.~W.}\ \bibnamefont {Bauer}},\ }\bibfield  {title}
  {\enquote {\bibinfo {title} {Magnon-polaron transport in magnetic
  insulators},}\ }\href {\doibase 10.1103/PhysRevB.95.144420} {\bibfield
  {journal} {\bibinfo  {journal} {Phys. Rev. B}\ }\textbf {\bibinfo {volume}
  {95}},\ \bibinfo {pages} {144420} (\bibinfo {year} {2017})}\BibitemShut
  {NoStop}%
\bibitem [{\citenamefont {Kikkawa}\ \emph {et~al.}(2016)\citenamefont
  {Kikkawa}, \citenamefont {Shen}, \citenamefont {Flebus}, \citenamefont
  {Duine}, \citenamefont {Uchida}, \citenamefont {Qiu}, \citenamefont {Bauer},\
  and\ \citenamefont {Saitoh}}]{Kikkawa2016}%
  \BibitemOpen
  \bibfield  {author} {\bibinfo {author} {\bibfnamefont {Takashi}\ \bibnamefont
  {Kikkawa}}, \bibinfo {author} {\bibfnamefont {Ka}~\bibnamefont {Shen}},
  \bibinfo {author} {\bibfnamefont {Benedetta}\ \bibnamefont {Flebus}},
  \bibinfo {author} {\bibfnamefont {Rembert~A.}\ \bibnamefont {Duine}},
  \bibinfo {author} {\bibfnamefont {Ken-ichi}\ \bibnamefont {Uchida}}, \bibinfo
  {author} {\bibfnamefont {Zhiyong}\ \bibnamefont {Qiu}}, \bibinfo {author}
  {\bibfnamefont {Gerrit E.~W.}\ \bibnamefont {Bauer}}, \ and\ \bibinfo
  {author} {\bibfnamefont {Eiji}\ \bibnamefont {Saitoh}},\ }\bibfield  {title}
  {\enquote {\bibinfo {title} {Magnon polarons in the spin seebeck effect},}\
  }\href {\doibase 10.1103/PhysRevLett.117.207203} {\bibfield  {journal}
  {\bibinfo  {journal} {Phys. Rev. Lett.}\ }\textbf {\bibinfo {volume} {117}},\
  \bibinfo {pages} {207203} (\bibinfo {year} {2016})}\BibitemShut {NoStop}%
\bibitem [{\citenamefont {Bozhko}\ \emph {et~al.}(2017)\citenamefont {Bozhko},
  \citenamefont {Clausen}, \citenamefont {Melkov}, \citenamefont {L'vov},
  \citenamefont {Pomyalov}, \citenamefont {Vasyuchka}, \citenamefont {Chumak},
  \citenamefont {Hillebrands},\ and\ \citenamefont {Serga}}]{Bozhko2017}%
  \BibitemOpen
  \bibfield  {author} {\bibinfo {author} {\bibfnamefont {Dmytro~A.}\
  \bibnamefont {Bozhko}}, \bibinfo {author} {\bibfnamefont {Peter}\
  \bibnamefont {Clausen}}, \bibinfo {author} {\bibfnamefont {Gennadii~A.}\
  \bibnamefont {Melkov}}, \bibinfo {author} {\bibfnamefont {Victor~S.}\
  \bibnamefont {L'vov}}, \bibinfo {author} {\bibfnamefont {Anna}\ \bibnamefont
  {Pomyalov}}, \bibinfo {author} {\bibfnamefont {Vitaliy~I.}\ \bibnamefont
  {Vasyuchka}}, \bibinfo {author} {\bibfnamefont {Andrii~V.}\ \bibnamefont
  {Chumak}}, \bibinfo {author} {\bibfnamefont {Burkard}\ \bibnamefont
  {Hillebrands}}, \ and\ \bibinfo {author} {\bibfnamefont {Alexander~A.}\
  \bibnamefont {Serga}},\ }\bibfield  {title} {\enquote {\bibinfo {title}
  {Bottleneck accumulation of hybrid magnetoelastic bosons},}\ }\href {\doibase
  10.1103/PhysRevLett.118.237201} {\bibfield  {journal} {\bibinfo  {journal}
  {Phys. Rev. Lett.}\ }\textbf {\bibinfo {volume} {118}},\ \bibinfo {pages}
  {237201} (\bibinfo {year} {2017})}\BibitemShut {NoStop}%
\bibitem [{\citenamefont {R\"uckriegel}\ \emph {et~al.}(2014)\citenamefont
  {R\"uckriegel}, \citenamefont {Kopietz}, \citenamefont {Bozhko},
  \citenamefont {Serga},\ and\ \citenamefont {Hillebrands}}]{Ruckriegel2014}%
  \BibitemOpen
  \bibfield  {author} {\bibinfo {author} {\bibfnamefont {Andreas}\ \bibnamefont
  {R\"uckriegel}}, \bibinfo {author} {\bibfnamefont {Peter}\ \bibnamefont
  {Kopietz}}, \bibinfo {author} {\bibfnamefont {Dmytro~A.}\ \bibnamefont
  {Bozhko}}, \bibinfo {author} {\bibfnamefont {Alexander~A.}\ \bibnamefont
  {Serga}}, \ and\ \bibinfo {author} {\bibfnamefont {Burkard}\ \bibnamefont
  {Hillebrands}},\ }\bibfield  {title} {\enquote {\bibinfo {title}
  {Magnetoelastic modes and lifetime of magnons in thin yttrium iron garnet
  films},}\ }\href {\doibase 10.1103/PhysRevB.89.184413} {\bibfield  {journal}
  {\bibinfo  {journal} {Phys. Rev. B}\ }\textbf {\bibinfo {volume} {89}},\
  \bibinfo {pages} {184413} (\bibinfo {year} {2014})}\BibitemShut {NoStop}%
\bibitem [{\citenamefont {Li}\ \emph {et~al.}(2012)\citenamefont {Li},
  \citenamefont {Ren}, \citenamefont {Wang}, \citenamefont {Zhang},
  \citenamefont {H\"anggi},\ and\ \citenamefont {Li}}]{RevModPhys.84.1045}%
  \BibitemOpen
  \bibfield  {author} {\bibinfo {author} {\bibfnamefont {Nianbei}\ \bibnamefont
  {Li}}, \bibinfo {author} {\bibfnamefont {Jie}\ \bibnamefont {Ren}}, \bibinfo
  {author} {\bibfnamefont {Lei}\ \bibnamefont {Wang}}, \bibinfo {author}
  {\bibfnamefont {Gang}\ \bibnamefont {Zhang}}, \bibinfo {author}
  {\bibfnamefont {Peter}\ \bibnamefont {H\"anggi}}, \ and\ \bibinfo {author}
  {\bibfnamefont {Baowen}\ \bibnamefont {Li}},\ }\bibfield  {title} {\enquote
  {\bibinfo {title} {Colloqium: Phononics: Manipulating heat flow with
  electronic analogs and beyond},}\ }\href {\doibase
  10.1103/RevModPhys.84.1045} {\bibfield  {journal} {\bibinfo  {journal} {Rev.
  Mod. Phys.}\ }\textbf {\bibinfo {volume} {84}},\ \bibinfo {pages}
  {1045--1066} (\bibinfo {year} {2012})}\BibitemShut {NoStop}%
\bibitem [{\citenamefont {Maldovan}(2013)}]{Maldovan_Nature_2013}%
  \BibitemOpen
  \bibfield  {author} {\bibinfo {author} {\bibfnamefont {Martin}\ \bibnamefont
  {Maldovan}},\ }\bibfield  {title} {\enquote {\bibinfo {title} {Sound and heat
  revolutions in phononics},}\ }\href {\doibase 10.1038/nature12608} {\bibfield
   {journal} {\bibinfo  {journal} {Nature}\ }\textbf {\bibinfo {volume}
  {503}},\ \bibinfo {pages} {209} (\bibinfo {year} {2013})}\BibitemShut
  {NoStop}%
\bibitem [{\citenamefont {Liu}\ \emph {et~al.}(2017{\natexlab{b}})\citenamefont
  {Liu}, \citenamefont {Xu}, \citenamefont {Zhang},\ and\ \citenamefont
  {Duan}}]{Zhang_Topo_Phonon_2016}%
  \BibitemOpen
  \bibfield  {author} {\bibinfo {author} {\bibfnamefont {Yizhou}\ \bibnamefont
  {Liu}}, \bibinfo {author} {\bibfnamefont {Yong}\ \bibnamefont {Xu}}, \bibinfo
  {author} {\bibfnamefont {Shou-Cheng}\ \bibnamefont {Zhang}}, \ and\ \bibinfo
  {author} {\bibfnamefont {Wenhui}\ \bibnamefont {Duan}},\ }\bibfield  {title}
  {\enquote {\bibinfo {title} {Model for topological phononics and phonon
  diode},}\ }\href {\doibase 10.1103/PhysRevB.96.064106} {\bibfield  {journal}
  {\bibinfo  {journal} {Phys. Rev. B}\ }\textbf {\bibinfo {volume} {96}},\
  \bibinfo {pages} {064106} (\bibinfo {year} {2017}{\natexlab{b}})}\BibitemShut
  {NoStop}%
\bibitem [{Note1()}]{Note1}%
  \BibitemOpen
  \bibinfo {note} {The demagnetization energy is disregarded since it only
  causes minor shifts in the dispersion ~\cite {Kamra2017}.}\BibitemShut
  {Stop}%
\bibitem [{\citenamefont {Scully}\ and\ \citenamefont
  {Zubairy}(1997)}]{Scully1997_Quantum}%
  \BibitemOpen
  \bibfield  {author} {\bibinfo {author} {\bibfnamefont {Marlan~O.}\
  \bibnamefont {Scully}}\ and\ \bibinfo {author} {\bibfnamefont {M.~Suhail}\
  \bibnamefont {Zubairy}},\ }\href@noop {} {\emph {\bibinfo {title} {Quantum
  optics}}}\ (\bibinfo  {publisher} {Cambridge University Press},\ \bibinfo
  {address} {Cambridge},\ \bibinfo {year} {1997})\BibitemShut {NoStop}%
\bibitem [{\citenamefont {Nakata}\ \emph {et~al.}(2017)\citenamefont {Nakata},
  \citenamefont {Klinovaja},\ and\ \citenamefont {Loss}}]{Nakata2017_Magnonic}%
  \BibitemOpen
  \bibfield  {author} {\bibinfo {author} {\bibfnamefont {Kouki}\ \bibnamefont
  {Nakata}}, \bibinfo {author} {\bibfnamefont {Jelena}\ \bibnamefont
  {Klinovaja}}, \ and\ \bibinfo {author} {\bibfnamefont {Daniel}\ \bibnamefont
  {Loss}},\ }\bibfield  {title} {\enquote {\bibinfo {title} {Magnonic quantum
  hall effect and wiedemann-franz law},}\ }\href {\doibase
  10.1103/PhysRevB.95.125429} {\bibfield  {journal} {\bibinfo  {journal} {Phys.
  Rev. B}\ }\textbf {\bibinfo {volume} {95}},\ \bibinfo {pages} {125429}
  (\bibinfo {year} {2017})}\BibitemShut {NoStop}%
\bibitem [{sup()}]{suppMat}%
  \BibitemOpen
  \href@noop {} {}\bibinfo {note} {See supplemental material.}\BibitemShut
  {Stop}%
\bibitem [{\citenamefont {Kittel}(1987)}]{Kittel1987}%
  \BibitemOpen
  \bibfield  {author} {\bibinfo {author} {\bibfnamefont {C.}~\bibnamefont
  {Kittel}},\ }\href@noop {} {\emph {\bibinfo {title} {Quantum Theory of
  Solids}}}\ (\bibinfo  {publisher} {John Wiley \& Sons},\ \bibinfo {year}
  {1987})\BibitemShut {NoStop}%
\bibitem [{\citenamefont {Kino}(1987)}]{Kino1987}%
  \BibitemOpen
  \bibfield  {author} {\bibinfo {author} {\bibfnamefont {G.S.}\ \bibnamefont
  {Kino}},\ }\href {https://books.google.no/books?id=hcsYAQAAIAAJ} {\emph
  {\bibinfo {title} {Acoustic Waves: Devices, Imaging, and Analog Signal
  Processing}}},\ Prentice-Hall Contemporary Topics in Accounting Series\
  (\bibinfo  {publisher} {Prentice-Hall},\ \bibinfo {year} {1987})\BibitemShut
  {NoStop}%
\bibitem [{\citenamefont {Maradudin}\ and\ \citenamefont
  {Stegeman}(1991)}]{Maradudin1991}%
  \BibitemOpen
  \bibfield  {author} {\bibinfo {author} {\bibfnamefont {A.~A.}\ \bibnamefont
  {Maradudin}}\ and\ \bibinfo {author} {\bibfnamefont {G.~I}\ \bibnamefont
  {Stegeman}},\ }\bibfield  {title} {\enquote {\bibinfo {title} {Surface
  acoustic waves},}\ }in\ \href@noop {} {\emph {\bibinfo {booktitle} {Surface
  phonons}}},\ \bibinfo {editor} {edited by\ \bibinfo {editor} {\bibfnamefont
  {F.~W. de~Wette}\ \bibnamefont {W.~Kress}}}\ (\bibinfo  {publisher}
  {Springer-Verlag},\ \bibinfo {address} {Berlin-Heidelberg},\ \bibinfo {year}
  {1991})\ Chap.~\bibinfo {chapter} {2}, pp.\ \bibinfo {pages}
  {5--36}\BibitemShut {NoStop}%
\bibitem [{\citenamefont {Bernevig}\ and\ \citenamefont
  {Hughes}()}]{Bernevig2013}%
  \BibitemOpen
  \bibfield  {author} {\bibinfo {author} {\bibfnamefont {B.~Andrei}\
  \bibnamefont {Bernevig}}\ and\ \bibinfo {author} {\bibfnamefont {Taylor~L.}\
  \bibnamefont {Hughes}},\ }\href@noop {} {\emph {\bibinfo {title} {Topological
  insulators and topological superconductors}}}\ (\bibinfo  {publisher}
  {Princeton University Press})\BibitemShut {NoStop}%
\bibitem [{\citenamefont {Ruello}\ and\ \citenamefont
  {Gusev}(2015)}]{Ruello2015}%
  \BibitemOpen
  \bibfield  {author} {\bibinfo {author} {\bibfnamefont {Pascal}\ \bibnamefont
  {Ruello}}\ and\ \bibinfo {author} {\bibfnamefont {Vitalyi~E}\ \bibnamefont
  {Gusev}},\ }\bibfield  {title} {\enquote {\bibinfo {title} {{Physical
  mechanisms of coherent acoustic phonons generation by ultrafast laser
  action}},}\ }\href {\doibase https://doi.org/10.1016/j.ultras.2014.06.004}
  {\bibfield  {journal} {\bibinfo  {journal} {Ultrasonics}\ }\textbf {\bibinfo
  {volume} {56}},\ \bibinfo {pages} {21--35} (\bibinfo {year}
  {2015})}\BibitemShut {NoStop}%
\bibitem [{\citenamefont {Liu}\ \emph {et~al.}(2017{\natexlab{c}})\citenamefont
  {Liu}, \citenamefont {Cornelissen}, \citenamefont {Shan}, \citenamefont
  {Kuschel},\ and\ \citenamefont {van Wees}}]{Liu2017}%
  \BibitemOpen
  \bibfield  {author} {\bibinfo {author} {\bibfnamefont {J.}~\bibnamefont
  {Liu}}, \bibinfo {author} {\bibfnamefont {L.~J.}\ \bibnamefont
  {Cornelissen}}, \bibinfo {author} {\bibfnamefont {J.}~\bibnamefont {Shan}},
  \bibinfo {author} {\bibfnamefont {T.}~\bibnamefont {Kuschel}}, \ and\
  \bibinfo {author} {\bibfnamefont {B.~J.}\ \bibnamefont {van Wees}},\
  }\bibfield  {title} {\enquote {\bibinfo {title} {Magnon planar hall effect
  and anisotropic magnetoresistance in a magnetic insulator},}\ }\href
  {\doibase 10.1103/PhysRevB.95.140402} {\bibfield  {journal} {\bibinfo
  {journal} {Phys. Rev. B}\ }\textbf {\bibinfo {volume} {95}},\ \bibinfo
  {pages} {140402} (\bibinfo {year} {2017}{\natexlab{c}})}\BibitemShut
  {NoStop}%
\bibitem [{Note2()}]{Note2}%
  \BibitemOpen
  \bibinfo {note} {The inset of Fig. \ref {fig_3}(b) shows two phonon modes.
  One is irrelevant since it is localized on the opposite edge.}\BibitemShut
  {Stop}%
\bibitem [{\citenamefont {Gowtham}\ \emph {et~al.}(2015)\citenamefont
  {Gowtham}, \citenamefont {Moriyama}, \citenamefont {Ralph},\ and\
  \citenamefont {Buhrman}}]{Gowtham2015}%
  \BibitemOpen
  \bibfield  {author} {\bibinfo {author} {\bibfnamefont {P.~G.}\ \bibnamefont
  {Gowtham}}, \bibinfo {author} {\bibfnamefont {T.}~\bibnamefont {Moriyama}},
  \bibinfo {author} {\bibfnamefont {D.~C.}\ \bibnamefont {Ralph}}, \ and\
  \bibinfo {author} {\bibfnamefont {R.~A.}\ \bibnamefont {Buhrman}},\
  }\bibfield  {title} {\enquote {\bibinfo {title} {Traveling surface spin-wave
  resonance spectroscopy using surface acoustic waves},}\ }\href {\doibase
  10.1063/1.4938390} {\bibfield  {journal} {\bibinfo  {journal} {Journal of
  Applied Physics}\ }\textbf {\bibinfo {volume} {118}},\ \bibinfo {pages}
  {233910} (\bibinfo {year} {2015})}\BibitemShut {NoStop}%
\bibitem [{\citenamefont {Datta}(1986)}]{Datta1986}%
  \BibitemOpen
  \bibfield  {author} {\bibinfo {author} {\bibfnamefont {S.}~\bibnamefont
  {Datta}},\ }\href {https://books.google.no/books?id=ng9TAAAAMAAJ} {\emph
  {\bibinfo {title} {Surface acoustic wave devices}}}\ (\bibinfo  {publisher}
  {Prentice-Hall},\ \bibinfo {year} {1986})\BibitemShut {NoStop}%
\bibitem [{\citenamefont {Mamishev}\ \emph {et~al.}(2004)\citenamefont
  {Mamishev}, \citenamefont {Sundara-Rajan}, \citenamefont {Yang},
  \citenamefont {Du},\ and\ \citenamefont {Zahn}}]{Mamishev2004}%
  \BibitemOpen
  \bibfield  {author} {\bibinfo {author} {\bibfnamefont {Alexander~V.}\
  \bibnamefont {Mamishev}}, \bibinfo {author} {\bibfnamefont {Kishore}\
  \bibnamefont {Sundara-Rajan}}, \bibinfo {author} {\bibfnamefont {Fumin}\
  \bibnamefont {Yang}}, \bibinfo {author} {\bibfnamefont {Yanqing}\
  \bibnamefont {Du}}, \ and\ \bibinfo {author} {\bibfnamefont {Markus}\
  \bibnamefont {Zahn}},\ }\bibfield  {title} {\enquote {\bibinfo {title}
  {Interdigital sensors and transducers},}\ }\href {\doibase
  10.1109/JPROC.2004.826603} {\bibfield  {journal} {\bibinfo  {journal}
  {Proceedings of the IEEE}\ }\textbf {\bibinfo {volume} {92}},\ \bibinfo
  {pages} {808} (\bibinfo {year} {2004})}\BibitemShut {NoStop}%
\end{thebibliography}%

\newpage

\begin{widetext}
\begin{center}
{\bf{Supplemental Material to the manuscript ``Chiral Phonon Transport Induced by Topological Magnons'' by}}\\
Even Thingstad, Akashdeep Kamra, Arne Brataas, Asle Sudb{\o}
\end{center}

\section{Rayleigh-like phonon edge modes}

\noindent To describe the phonons, as discussed in the main paper, we consider a force constant  model for out-of-plane phonon modes on the honeycomb lattice with only nearest neighbour interaction. This is described by the Hamiltonian 


\begin{equation}
H = \sum_i  \frac{p_i^2}{2m} + \frac{1}{2}  \sum_{\langle i,j\rangle} C (u_{i} - u_{j} )^2,
\end{equation}

\noindent where \(i\) and \(j\) are lattice site indices running over both the \(A\) and \(B\) sublattices of the honeycomb lattice.  \\

\noindent To investigate edge modes in the system, we consider a finite ribbon geometry with periodic boundary conditions in one direction, and with a finite number of unit cells in the other direction. The edges of such ribbons can mainly be of two types: zigzag and armchair. The lattice geometries of these cases are shown in Fig. \ref{fig_ribbonLattices}.  \\

\noindent To find the phonon energy spectrum for these lattice geometries, we introduce the partial Fourier transform of the lattice site deviations and momenta, which for the lattice site deviation takes the form

\begin{equation}
u_{x,y}^D= \frac{1}{\sqrt{N_x}} \sum_k u_{ky} \exp\left(i k \hat{x} \cdot \mathbf{r}_{n}^D \right) ,\\
\end{equation}

\noindent where \(u_{x,y}^D\) is the lattice site deviation on sublattice \(D\) in unit cell \((x,y)\),  \(\mathbf{r}_{x,y}^D\) is the corresponding equilibrium position, and \(N_x\) is the number of unit cells in the horizontal direction. The periodicity requirement \(u_{x,y}^D = u_{x+N_x, y}^D \) then gives \(k= 2\pi n/ N_x \lambda\), where \(\lambda\) is the periodicity of the lattice in the horizontal direction and \(n\) is an integer. For the zigzag edge ribbon, \(\lambda = \sqrt{3} d\), while \(\lambda = 3d\) for the armchair edge ribbon. This determines the size of the Brillouin zone. \\

\noindent Introducing

\begin{equation}
u_k^\dagger = (u_{-k1}^A,\; u_{-k1}^B,\; u_{-k2}^A, \; u_{-k2}^B, \;\dots \;, \;u_{-k N_y}^A, \; u_{-k N_y}^B )
\end{equation}

\noindent with similar notation for the momentum, the phonon Hamiltonian can be written on the form 

\begin{equation}
H = \frac{1}{2m} \sum_k p_k^\dagger p_k + \frac{C}{2} \sum_k u_k^\dagger M_k u_k,
\end{equation}

\noindent with a matrix \(M_k\) coupling the deviations on the various sublattices and neighbouring unit cell layers. This Hamiltonian is diagonalized through a unitary transform of the deviations and momenta followed by introducing phonon creation and annihilation operators \(c_k^\dagger\) and \(c_k\) \cite{Kittel1987}. The excitation spectrum is then given by the phonon frequencies \(\omega_{kn}\), where
\(\omega_{kn}^2 =  (C/m) \lambda_{kn}\)
and \(\{ \lambda_{kn} \}\) are the eigenvalues of \(M_k\). \\

\begin{figure}[tbp]
\centering
\subfigure[\; Zigzag edge ribbon.]   {\includegraphics[width=0.4\textwidth]{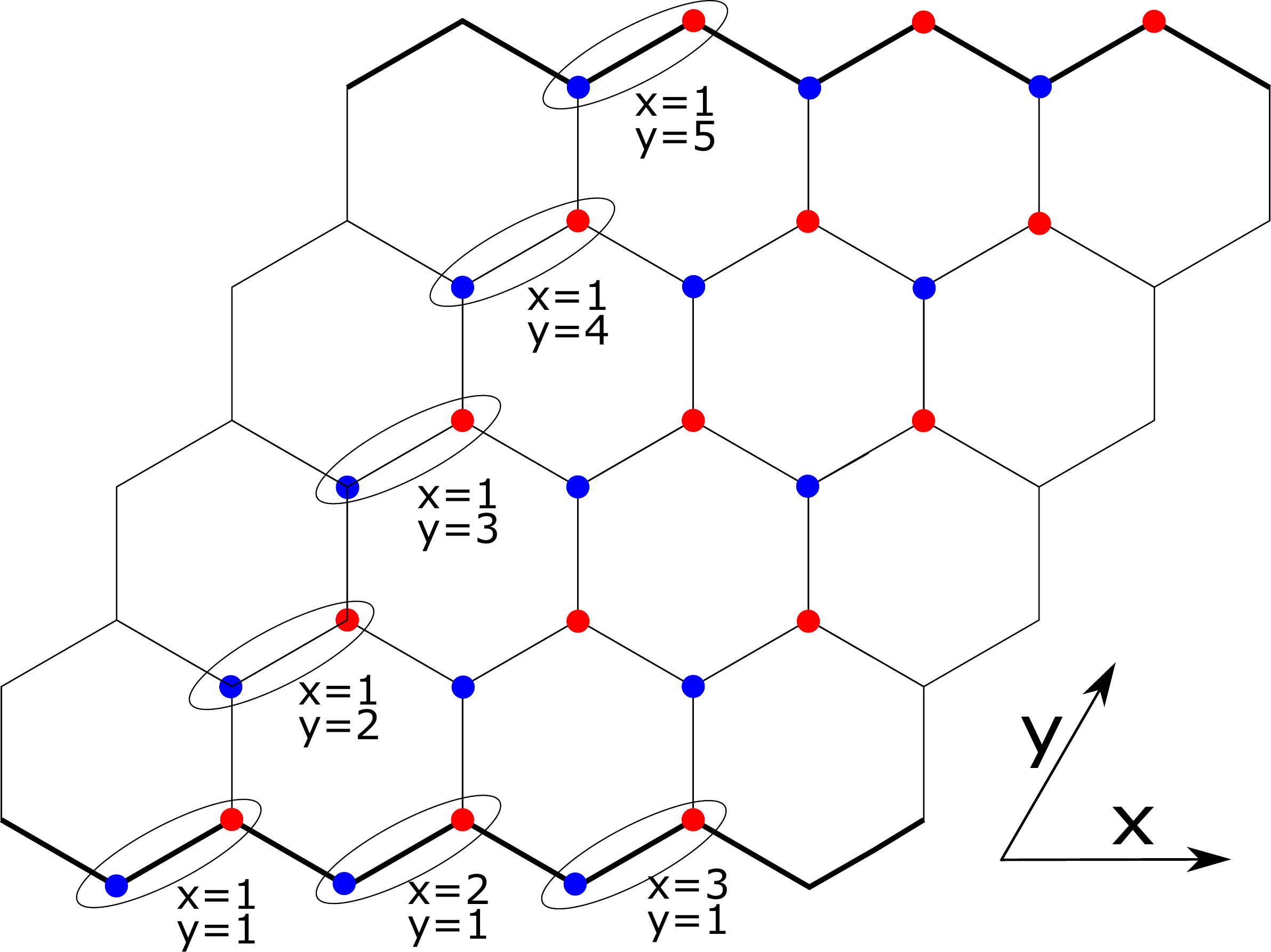}}\label{fig_lattice}
\subfigure[\; Armchair edge ribbon] {\includegraphics[width=0.4\textwidth]{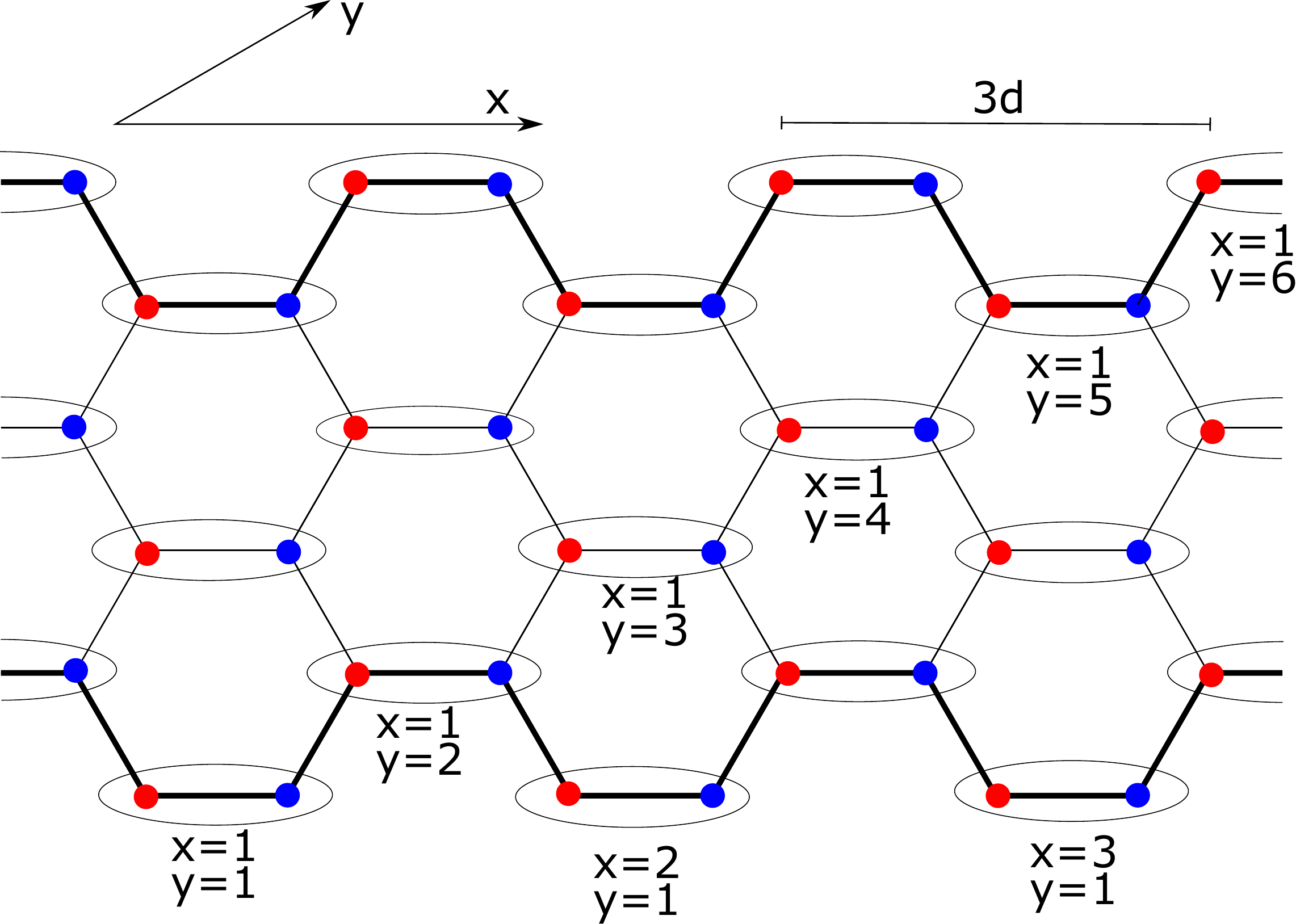}}\label{fig:1b}
\caption{\sf Lattice geometries for the zigzag- and armchair edge ribbons, including unit cell labelling. We assume periodic boundary conditions in the horizontal direction and a finite number of hexagon layers in the vertical direction. }
\label{fig_ribbonLattices}
\end{figure}

\noindent Following this procedure for the ribbon geometry with zigzag edges, we obtain the spectrum in Fig. \ref{fig_ribbonSpectraPhonon}(a). The upper and lower branches of the phonon spectrum meet at \(k_x d = 2\pi/3\sqrt{3}\) and   \(k_x d = 4\pi/3\sqrt{3}\), consistent with the result obtained by taking the 1-dimensional projection of the bulk bands. In Fig. \ref{fig_phononEigenstates_zigzag}, we plot the spatial profile of some selected eigenstates at quasimomentum \(k_x d = 2\pi/3\sqrt{3}\). All the modes are delocalized.  \\

\noindent Examining the armchair ribbon spectrum in Fig. \ref{fig_ribbonSpectraPhonon}(b),
one may notice that the modes marked with green arrows stand out from the rest.
If we were to compute the bulk spectrum and then perform a 1D projection, the two modes marked in green would not be found. We therefore conclude that they must originate from an edge effect. \\

\noindent This is confirmed by examining the spatial profile of the modes, as shown in Fig. \ref{fig_phononEdgeLocalization} for the modes between the upper and lower bulk phonon branches. The deviation amplitudes are finite on the outer armchair edges of the sample, and  exponentially decaying into the interior of the sample. The inset shows the decay length as function of the inverse quasimomentum, and demonstrates that \(\xi \propto 1/k_x\). This is perfectly analogous to the behaviour of so-called Rayleigh modes on the surface of a three-dimensional material \cite{Maradudin1991}. Our modes can therefore be characterized as one-dimensional analogs of Rayleigh modes. \\

\noindent From the above discussion, it follows that that while the armchair edges support edge modes, the zigzag edge does not. This is rooted in the fact that on the edge unit cells of the armchair ribbon, both atoms have 2 nearest neighbours. For the zigzag ribbon, one atom has 2 nearest neighbours, but the other has 3.  Vibrations are therefore easier to excite on the edges of the armchair ribbon.

\begin{figure}[tbp]
\centering
\subfigure[\; Zigzag edge ribbon]   {\includegraphics[width=0.45\textwidth]{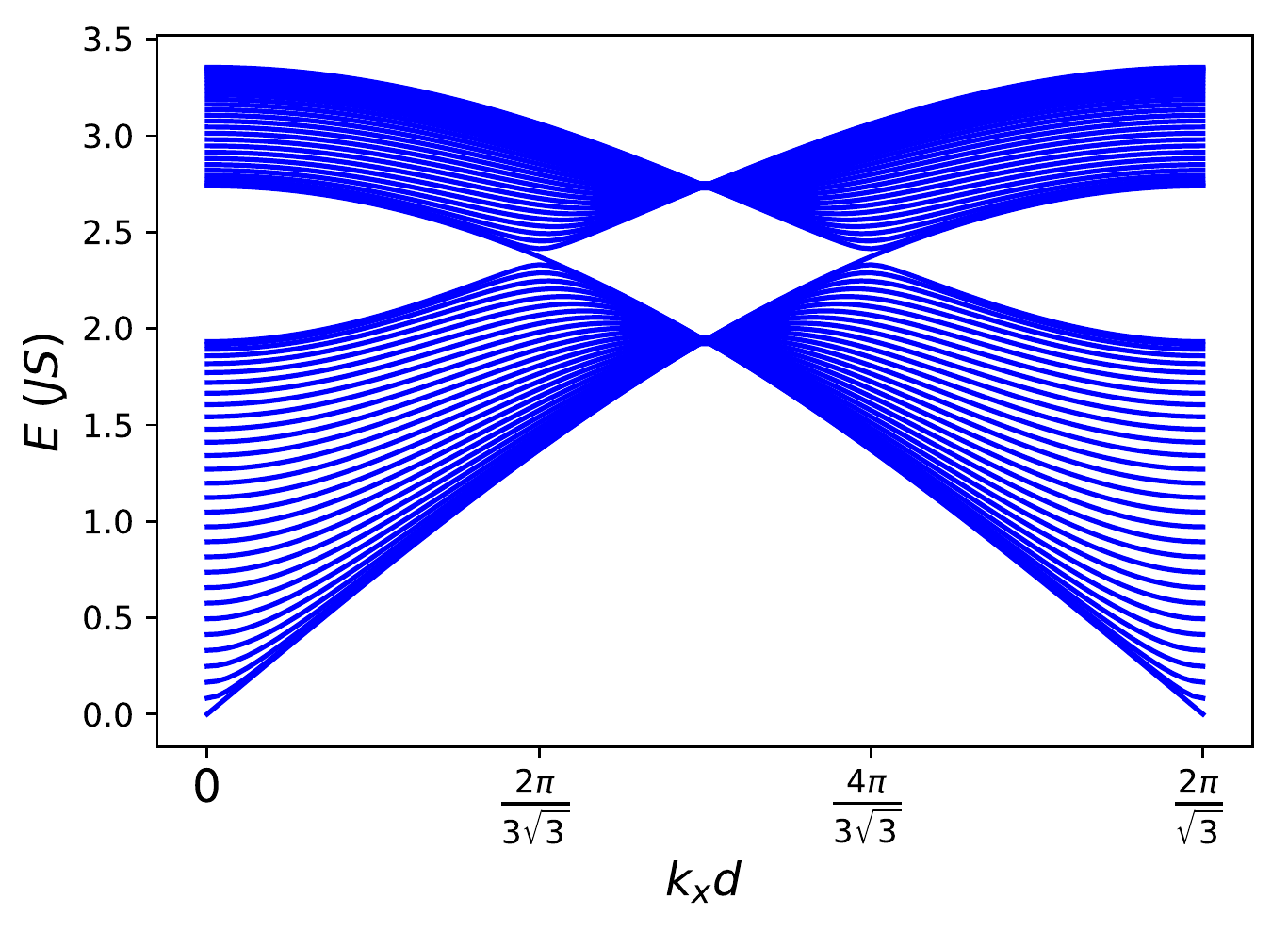}}\label{fig_ribbonSpectraPhonon_zigzag}
\subfigure[\; Armchair edge ribbon] {\includegraphics[width=0.45\textwidth]{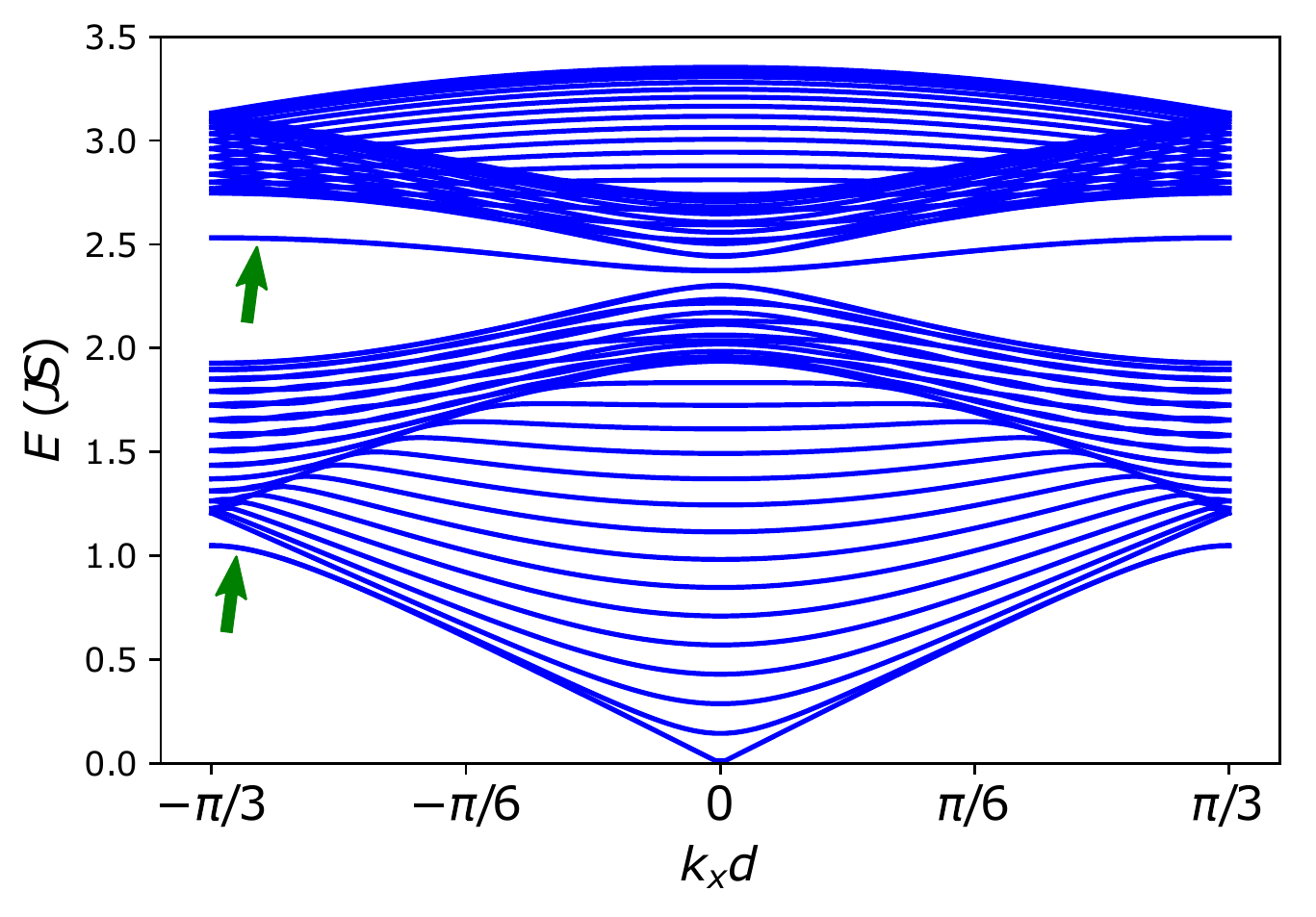}}\label{fig_ribbonSpectraPhonon_armchair}
\caption{\sf One-dimensional projection of the energy spectrum for phonon modes on the honeycomb ribbon geometry with zigzag and armchair edges for \(N_y=30\) unit cells in the vertical direction (see Fig. \ref{fig_ribbonLattices}). For the armchair ribbon, in addition to the bulk modes, there are edge modes marked with green arrows.  }
\label{fig_ribbonSpectraPhonon}
\end{figure}

\begin{figure}
\centering
\includegraphics[width=0.6\textwidth]{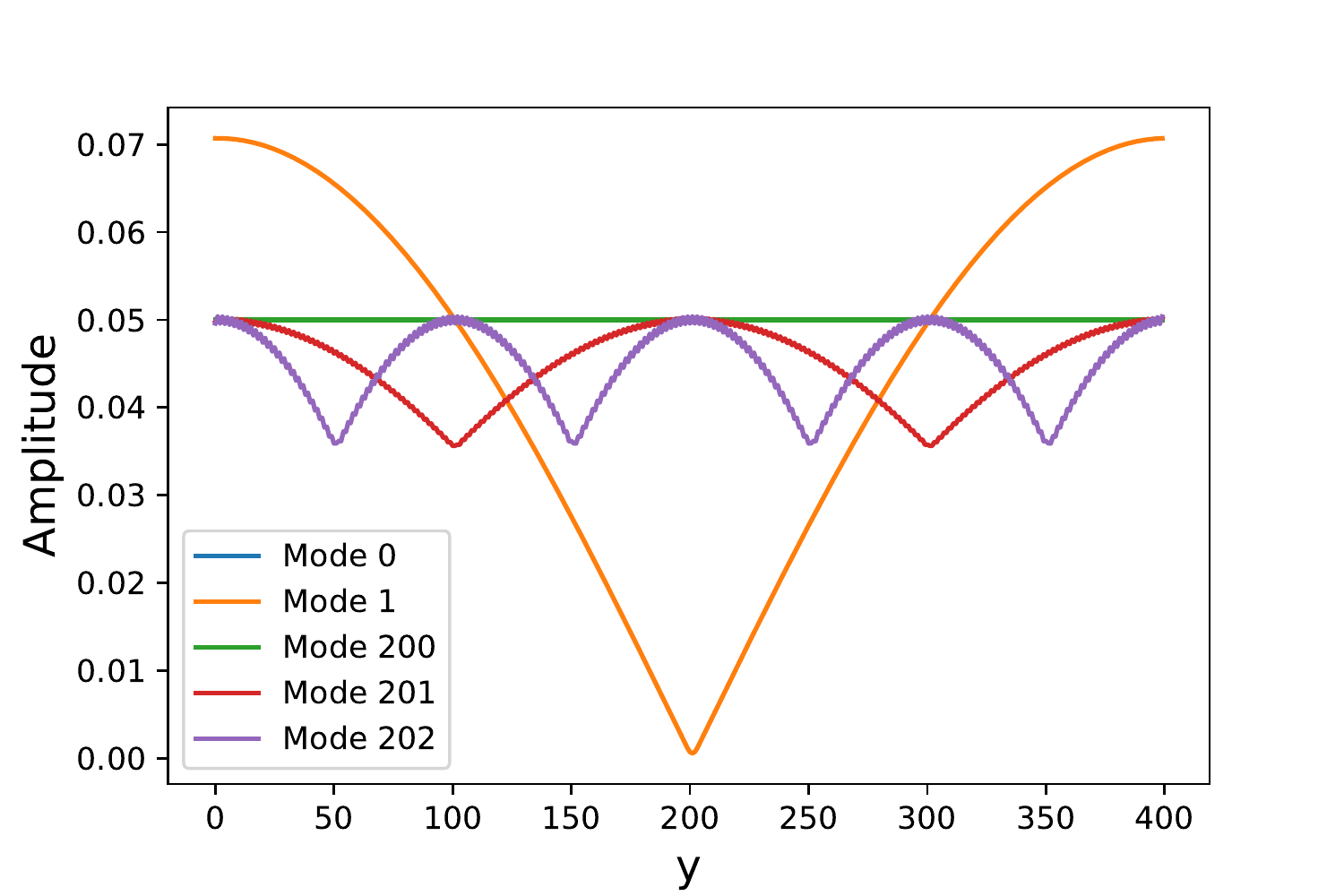}
\caption{ \sf Deviation amplitudes for selected zigzag ribbon phonon modes at \(k_x d = 2 \pi /3\sqrt{3}\) for \(N_y=200\) unit cells in the vertical direction. Fast oscillations have been averaged out. All modes are delocalized. }
\label{fig_phononEigenstates_zigzag}
\end{figure}


\begin{figure}
\centering
\includegraphics[width=0.8\textwidth]{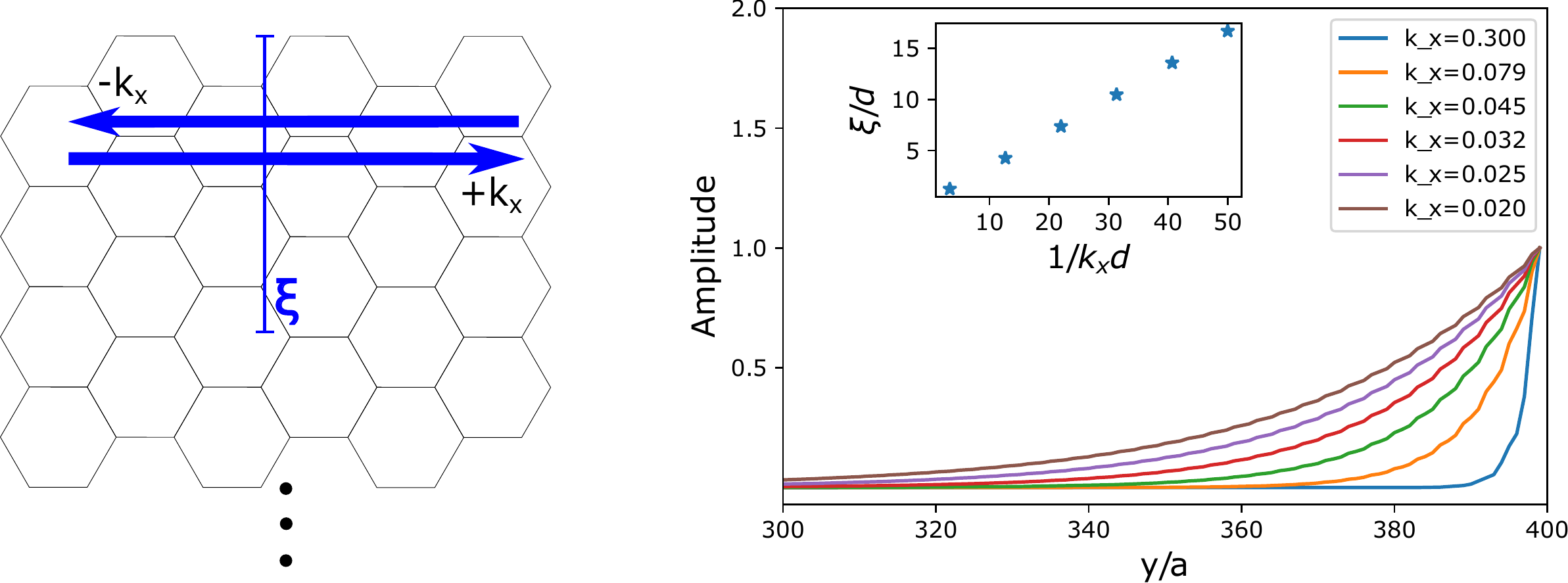}
\caption{\sf (a) Rayleigh mode schematic. At a given armchair edge, two Rayleigh-like modes propagate along the edge with quasimomenta \(\pm k_x\). The modes are localized within a distance \(\xi(k_x)\) from the edge. (b) Deviation amplitudes for the Rayleigh-like edge modes as function of the vertical position \(y\) in units of the Bravais lattice constant  \(a\) for different momenta \(k_x\). The amplitudes are normalized to the value on the edge for easier comparison. The inset shows the localization length \(\xi(k_x)\) as function of the inverse quasimomentum, demonstrating that \(\xi \propto 1/k_x\), consistent with the behaviour expected from ordinary Rayleigh modes. }
\label{fig_phononEdgeLocalization}
\end{figure}

\section{Coupled magnetoelastic modes in Zigzag ribbon}

\noindent To compute the excitation spectrum for the model with coupled magnon and phonon modes, we first calculate the phonon and magnon edge modes for the uncoupled model. The phonon modes were discussed in the previous section, and we refer to the literature for the magnon spectrum \cite{Bernevig2013}.  Expressing the magneto-elastic coupling in terms of these eigenmodes and diagonalizing the resulting matrix, we obtain the excitation spectra.  \\

\noindent For the zigzag edge ribbon, the spectrum is shown in Fig. \ref{fig_ribbonSpectra}. All phonon modes are delocalized. In the inset, we show the hybridization of the chiral edge magnon mode with some of these delocalized modes. The armchair ribbon spectrum has already been discussed in the main text. 

\begin{figure}
\centering
\includegraphics[width=0.6\textwidth]{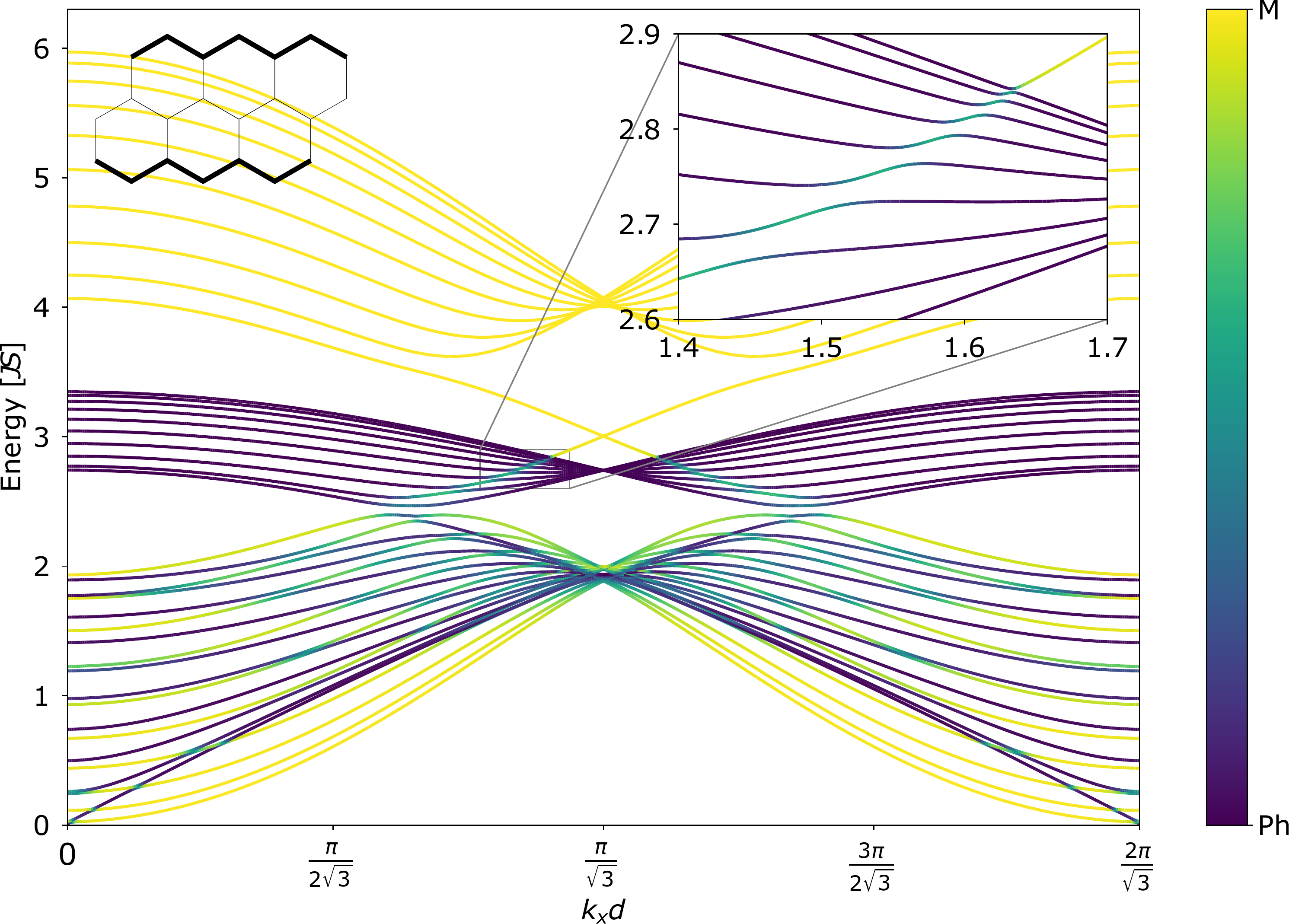}
\caption{\sf Energy spectrum for the coupled magnon and phonon modes of the ribbon geometry with zigzag edge as function of quasimomentum \(k_x\). The magnon (yellow) and phonon (purple) content of each mode is indicated with color. }
\label{fig_ribbonSpectra}
\end{figure}

\section{Interdigital elastic transducers}

\subsection{General principles and qualitative description}

\begin{figure}
    \centering
    \includegraphics[width=110mm]{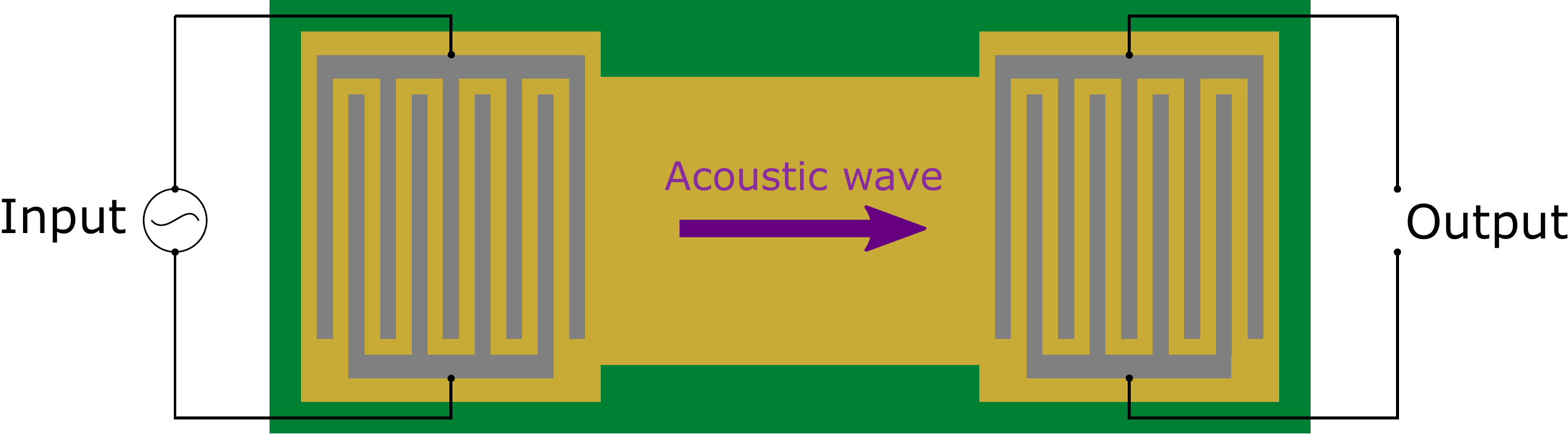}
    \caption{Schematic depiction of an interdigital transducer employed to excite Rayleigh waves via an ac voltage on the left hand side. The same structure converts the acoustic waves back to ac voltage on the right and enables their detection. The metallic electrodes are lithographically patterned on top of a piezoelectric material, such as Lithium Niobate, into the depicted comb structure. }
    \label{fig:IDT_schematic}
\end{figure}

\noindent The interdigital transducer~\cite{Kino1987,Mamishev2004} (IDT) consists of two metallic electrodes with a series of sections, called fingers, which are patterned into a comb-like structure on top of a piezoelectric material (Fig. \ref{fig:IDT_schematic}). When a voltage is applied across the two electrodes, it creates a pattern of alternating charges on adjacent fingers via the capacitive effect. Via the constitutive properties of the piezoelectric material, this results in a pattern of alternating strains. An applied ac voltage with angular frequency $\omega$ thus excites acoustic waves at the same frequency in the piezoelectric material. The wavelength is determined by the corresponding dispersion relation $\omega = c k$ ~\footnote{For simplicity, and as a concrete example, we consider the linear dispersion of the Rayleigh waves typical of a conventional IDT system. The technique however works for an arbitrary dispersion $\omega(k)$.}, where $c$ is the speed of sound in the material and $k$ is the wavenumber. If the ensuing wavelength $\lambda = 2 \pi / k$ is equal to the spacing between the adjacent fingers belonging to the same electrode, the acoustic signal interferes constructively and the excitation efficiency is high. If there is a mismatch between the finger spacing and the wavelength excited at the applied ac voltage frequency, the acoustic waves tend to cancel each other and excitation efficiency is low. With an increasing number $N$ of fingers, the reinforcement or cancellation effect is stronger and the excitation resonances become sharper. Thus, the operation principle of an IDT is similar to that of a Bragg grating. Then, it is easy to understand that peaks in excitation are observed at multiple frequencies (and wavelengths) corresponding to the finger spacing being multiples of the acoustic wavelength. The fundamental peak is the strongest and subsequent overtones are progressively weaker as demonstrated by the frequency transfer characteristics discussed below. \\

\noindent Conventionally, IDTs have been employed in applications such as analog filters, and their desired operation frequency range has been from MHz to several tens of GHz~\cite{Kino1987}. A typical piezoelectric material employed is Lithium Niobate with a Rayleigh wave speed of 3.3 km$/$s. Thus, the fundamental peak corresponding to a center frequency of 1 GHz requires finger spacing of around 1 $\mu$m, which could easily be achieved via photolithography techniques. With contemporary electron-beam lithography techniques, a finger spacing of several tens of nanometers is readily possible, thus allowing a fundamental frequency of tens of GHz. Employing higher overtones allows pushing the operation frequency to several tens of GHz, and is predominantly limited by the driving electronics~\cite{Mamishev2004}. Due to the purview of their conventional applications, attempts to achieve higher frequencies have been limited. With recent advances in ultrafast lasers, several conventional methods have been adapted to achieve coherent phonon generation in the THz regime~\cite{Ruello2015}. Thus, the operation range of the proposed method is estimated to be rather wide with up to hundreds of GHz in frequency and tens of nanometers in wavelength. The wavenumber selectivity can also be increased, in principle to arbitrary values, by using a large number of fingers. Combined with the tunability of the exact magnon-phonon anticrossing point (via an applied field, for example) across a broad range of frequencies and wavevectors, the proposed experimental method is well within the range of the contemporary state-of-the-art technology. \\

\subsection{Frequency resolved acoustic output}

\begin{figure}
    \centering
    \includegraphics[width=60mm]{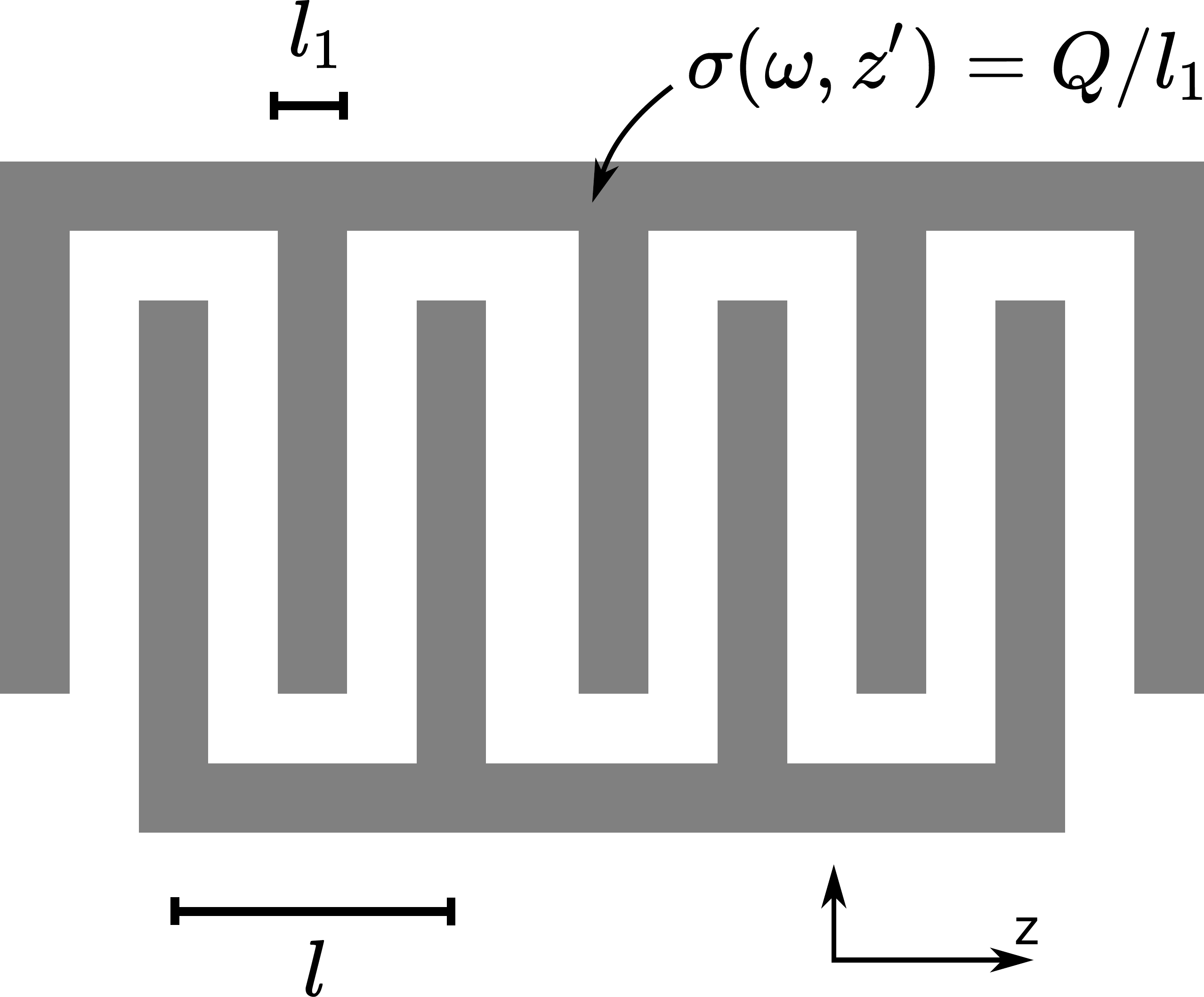}
    \caption{Modeling the acoustic output of an interdigital transducer within the delta function model. An applied voltage generates charges on the metallic electrodes via the capacitive effect. The accumulated charge (or equivalently electric field) is converted into strain via the piezoelectric effect.}
    \label{fig:IDT_model}
\end{figure}

\noindent To supplement the above qualitative discussion of the operating principles of an IDT, we now discuss its frequency resolved excitation efficiency within the so-called delta function model (Fig. \ref{fig:IDT_model}). This model assumes that the charge accumulated on each finger is distributed uniformly and that the acoustic output is a linear superposition of the strain produced by the full charge distribution. In evaluating the strain (and thus the acoustic output) at a given point due to different fingers, the phase difference due to wave propagation from the fingers needs to be accounted adequately. The amplitude of the excited acoustic wave $A(z,\omega)$ at a position $z$ is thus given by
\begin{align}
    A(z,\omega) = & \alpha \int \sigma(\omega,z^\prime) e^{-i k (z-z^\prime)} d z^\prime,
\end{align}
where $\alpha$ is the charge-strain coupling factor of the piezoelectric material and $\sigma(\omega,z^\prime)$ is the charge density accumulated at position $z^\prime$. The acoustic output of the IDT is thus simply the Fourier transform of the accumulated charge density. The square-wave like pattern of the accumulated charge on the comb-like structure thus suggests a sinc function response. \\

\noindent Referring the readers to detailed derivations and discussion in Ref. \onlinecite{Kino1987}, we simply present the key result here. For an $N$-finger comb, the overall acoustic amplitude outside the IDT region becomes
\begin{align}
    A(\omega,z) = & i \alpha Q \ \frac{\sin \left( k N l /2 \right)}{\cos \left( k l / 4 \right)} \mathrm{sinc} \left( \frac{l_1}{\lambda} \right) e^{i k (N-1)l / 2} \ e^{i(\omega t - k z)}, 
\end{align}
where $Q$ is the charge on a single finger, $l$ and $l_1$ are defined in Fig. \ref{fig:IDT_model}, and sinc$(x) = \sin(\pi x)/(\pi x)$. The equation above fully describes all the design characteristics of the device. Its further analysis shows that the center or fundamental frequency $\omega_0$ is determined by the condition $k_{0}l = 2 \pi$ (Note that $\omega_0 = c k_0$), while the bandwidth between the zeros in the response is given by $\Delta \omega/ \omega_0 = 2/ N$, in consistence with the qualitative discussion above.

\end{widetext}


\end{document}